\documentclass[
reprint,
superscriptaddress,
showpacs,
preprintnumbers,
nofootinbib,
nobibnotes,
amsmath,
amssymb, 
aps,
prd,
floatfix
]{revtex4-2}

\usepackage{graphicx}
\usepackage{dcolumn}
\usepackage{bm}
\usepackage{graphicx}
\usepackage{subfig}
\usepackage{amsmath}
\usepackage{siunitx}
\usepackage{xcolor}
\usepackage{hyperref}

\DeclareSIUnit \s {\second}
\DeclareSIUnit \ns {\nano\second}
\DeclareSIUnit \mus {\micro\second}
\DeclareSIUnit \ms {\milli\second}
\DeclareSIUnit \MB {\mega\byte}
\DeclareSIUnit \GB {\giga\byte}
\DeclareSIUnit \TB {\tera\byte}
\DeclareSIUnit \PB {\peta\byte}
\DeclareSIUnit \Mbps {\mega\bit/\s}
\DeclareSIUnit \Gbps {\giga\bit/\s}
\DeclareSIUnit \Tbps {\tera\bit/\s}
\DeclareSIUnit \Pbps {\peta\bit/\s}
\DeclareSIUnit \kton {\kilo\tonne} 
\DeclareSIUnit \kt {\kilo\tonne}
\DeclareSIUnit \Mt {\mega\tonne}
\DeclareSIUnit \eV {\electronvolt}
\DeclareSIUnit \keV {\kilo\electronvolt}
\DeclareSIUnit \MeV {\mega\electronvolt}
\DeclareSIUnit \GeV {\giga\electronvolt}
\DeclareSIUnit \PeV {\peta\electronvolt}
\DeclareSIUnit \EeV {\exa\electronvolt}
\DeclareSIUnit \m {\meter}
\DeclareSIUnit \cm {\centi\meter}
\DeclareSIUnit \in {\inchcommand}
\DeclareSIUnit \km {\kilo\meter}
\DeclareSIUnit \kV {\kilo\volt}
\DeclareSIUnit \kW {\kilo\watt}
\DeclareSIUnit \MW {\mega\watt}
\DeclareSIUnit \MHz {\mega\hertz}
\DeclareSIUnit \mrad {\milli\radian}
\DeclareSIUnit \year {years}
\DeclareSIUnit \POT {POT}
\DeclareSIUnit \sig {$\sigma$}
\DeclareSIUnit\parsec{pc}
\DeclareSIUnit\lightyear{ly}
\DeclareSIUnit\foot{ft}
\DeclareSIUnit\ft{ft}
\DeclareSIUnit \ppb{ppb}
\DeclareSIUnit \ppt{ppt}
\DeclareSIUnit \samples{S}
\DeclareSIUnit \pe{PE}
\DeclareSIUnit \sr{\steradian}
\DeclareSIUnit \Mtons{Mtons}

\begin{document}

\title{Sensitivity to Supernovae Average $\nu_x$ Temperature with Neutral Current Interactions in DUNE}

\author{Darcy A. Newmark}
\affiliation{Dept.~of Physics, Massachusetts Institute of Technology, Cambridge, MA 02139, USA}

\author{Austin Schneider}
\affiliation{Los Alamos Laboratory, Los Alamos, NM 87545, USA}
\affiliation{Dept.~of Physics, Massachusetts Institute of Technology, Cambridge, MA 02139, USA}

\date{\today}

\begin{abstract}
We explore a novel method for measuring the average temperature of the $\nu_x$ component in Type-II core-collapse supernovae.
By measuring neutral current incoherent neutrino-Argon interactions in DUNE we can obtain spectral information for the combination of all active neutrino species.
Combining this all-neutrino spectral information with detailed charged current measurements of the electron neutrino and electron anti-neutrino fluxes from DUNE and Hyper-Kamiokande, we can infer the average temperature for the remaining neutrino species in the $\nu_x$ component to within a factor two for most cases and to 30\% for a small range of average $\nu_x$ temperatures.
Due to the limited energy range of the emitted photons from incoherent neutral current interactions on Argon, the $\nu_x$ temperature reconstruction demonstrates a degeneracy in the one and two sigma credible regions.
Furthermore, while large uncertainties on the NC cross-section penalize this measurement, we examined the efficacy of constraining NC cross-section uncertainties on improving $\nu_x$ measurements.
We found that if additional measurements of B(M1$\uparrow$) 1$^+$ excited state transitions in Argon are able to reduce correlated cross section uncertainties from 15\% to 7\%, the size of the $1\sigma$ allowed regions for $T_{\nu_x}$ becomes sample size limited, and approaches the case where there are no uncertainties on the cross-section.
\end{abstract}

\maketitle

\section{Introduction\label{sec:intro}}
Type-II core-collapse supernovae events release approximately $10^{53}$ ergs of energy, carried away almost exclusively by neutrinos and antineutrinos of all flavors~\cite{Mirizzi:2015eza}.
These neutrinos travel nearly unimpeded to Earth, where they can be detected.
In 1987, two neutrino detectors on Earth measured 20 events within 13 seconds, which coincided with the release of energy from the collapse of SN1987A~\cite{Mirizzi:2015eza}.
If a core collapse supernovae event were to occur within the Galaxy today, the many operating neutrino detectors would provide precise measurements of neutrino and antineutrino energy spectra.
Such measurements are crucial for reconstructing the initial conditions of supernovae just before collapse, understanding their evolution, and modeling their collapse.

Large detectors such as the Deep Underground Neutrino Experiment (DUNE) and Hyper-Kamiokande (Hyper-K) will provide large sample measurements of $\nu_e$ and $\bar{\nu_e}$ charged current (CC) interactions, respectively, with excellent energy resolution~\cite{duneCC, Hyper-Kamiokande:2021frf}
However, the remaining neutrino species, summarized as $\nu_x$, are below threshold for CC interactions and can only be measured through neutral current (NC) interactions.
While there are many theoretical models of supernovae collapse, they vary greatly in the prediction and treatment of $\nu_x$~\cite{Raffelt:2001kv, Raffelt:2003en, Buras:2002wt}. 
With little existing experimental information and large uncertainties in the model space, measurements of NC neutrino interactions will be crucial for understanding the $\nu_x$  component of core collapse supernovae.
Using recently updated incoherent NC neutrino-nucleus cross-sections~\cite{Tornow:2022kmo}, we provide predictions and sensitivities for average $\nu_x$ temperature reconstructions.

\section{Measurement Strategy\label{sec:strategy}}
\begin{figure}
  \centering
  \includegraphics[width=\linewidth]{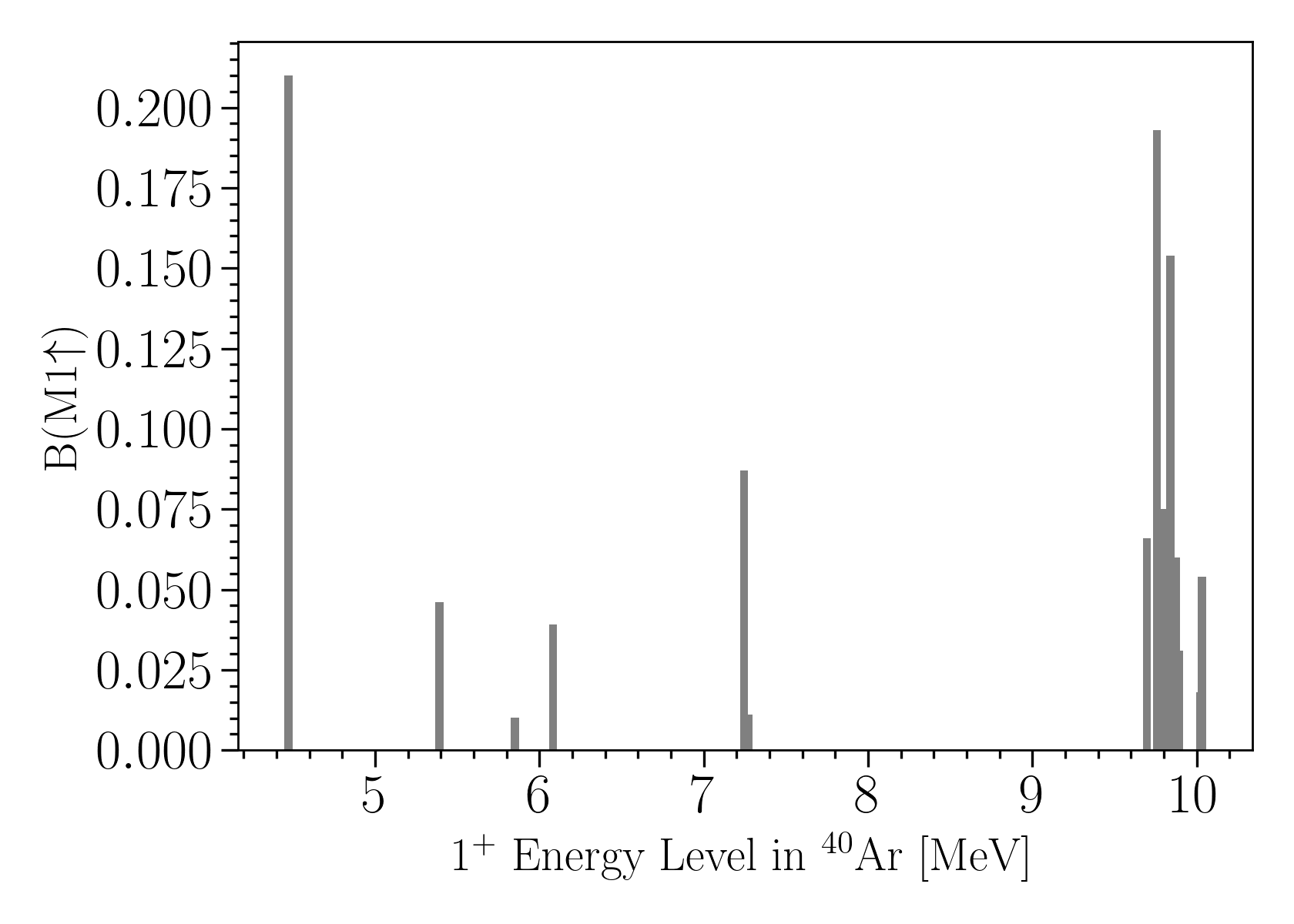}
  \caption{\textbf{B(M1$\uparrow$) neutral current $1^+$ transition strengths for $^{40}$Ar.} The B(M1$\uparrow$) value for each $1^+$ is shown as a function of the transition energy. Fourteen of these $1+$ transitions exist in total.}
  \label{fig:argon levels}
\end{figure}

In the incoherent neutral current interaction, Argon nucleons are excited via interaction with a neutrino. As the nucleon de-excites, $\gamma$-rays are emitted at specific energy levels.
The magnetic dipole strengths of these $1^+$ excited state transitions are displayed in Fig.~\ref{fig:argon levels}. 
These measurements, combined with shell model calculations, guide neutral current cross-section predictions and dictate experimental reconstruction of incident neutrino spectra. 

Neutral current neutrino-Argon interactions are difficult to measure due to the small cross-section of this channel and low energy at which the interactions occur~\cite{Tornow:2022kmo}.
For supernovae neutrino energies, the NC interaction cross-section is at least two orders of magnitude smaller than the corresponding CC cross-section, given by ~\cite{PhysRevC.80.055501}.
For this reason, projected measurements of supernova neutrinos tend to focus on CC channels.

While NC measurements are difficult to make compared to the CC channel, they provide the only source of information about the neutrino species of $\nu_x$.
Flavor non-specific information from supernovae neutrinos is crucial to understanding the neutrino flux emitted without uncertainties related to neutrino oscillations, matter effects, or neutrino self-interactions that occur within the supernovae.
Furthermore, measurements of all neutrino species will provide us with a more complete understanding of supernovae neutrinos.
Combining flavor non-specific NC measurements with $\nu_e$ and $\bar{\nu_e}$ CC measurements, we can indirectly probe the properties of the supernova $\nu_x$ flux.

Recent developments in detector technology will allow us to overcome this challenge with ultra-large detectors that are sensitive to these low energy NC interactions.
With \SI{40}{\kilo\tonne} of combined fiducial mass and energy resolution down to the \si\MeV{} scale, the DUNE Far Detector (FD) modules will be an ideal candidate to study the $\nu_x$ component, providing the largest sample of NC neutrino-Argon interactions~\cite{DUNE:2020ypp}.
DUNE will be able to separate CC from NC interactions using the distribution of deposited energy as a function of interaction length within the detector. 
The DUNE FD modules are planned to have a high efficiency trigger system down to the few \si\MeV{} threshold, with good energy resolution, and moderate spatial resolution~\cite{DUNE:2020ypp, flavio}. 
With these anticipated capabilities, DUNE will be able to resolve photon emissions from NC interaction in the energy range of interest for this analysis.

While this work is directly relevant to DUNE, SBND, ICARUS and other Liquid Argon (LAr) detectors, supernovae neutrino interaction measurements on other nuclei will provide further insights into the $\nu_x$ spectrum.
The Jiangmen Underground Neutrino Observatory (JUNO) experiment will be able to observe incoherent neutral current neutrino interactions on Carbon nuclei, with a predominant excitation state of \SI{15.1}\MeV~\cite{JUNO:2015sjr}.
A combined analysis of NC neutrino interactions on $^{40}$Ar and $^{12}$C would provide a robust and uniquely sensitive measurement of $\nu_x$ supernovae neutrinos.

In addition to NC measurements, DUNE and Hyper-K will provide leading constraints on the $\nu_e$ and $\bar{\nu_e}$ spectrum through CC measurements on liquid Argon and water, respectively. 
By incorporating DUNE CC, DUNE NC, and Hyper-K CC measurements in a combined analysis, we can make definitive statements about the $\nu_x$ component of the neutrino flux.
Although sample sizes for NC events are much smaller, and energy resolution is poorer, the information we obtain from NC interaction on Argon nuclei in DUNE provides enough discriminating power to reconstruct the average $\nu_x$ temperature.

\section{Methodology\label{sec:methods}}
Modelling the neutral current measurements requires three ingredients: the supernova neutrino flux as a function of neutrino temperature, the cross-section of the NC interaction, and DUNE's detector response.
Our choices for these three components of the calculation are described below.

\begin{figure}
 \includegraphics[width=\linewidth]{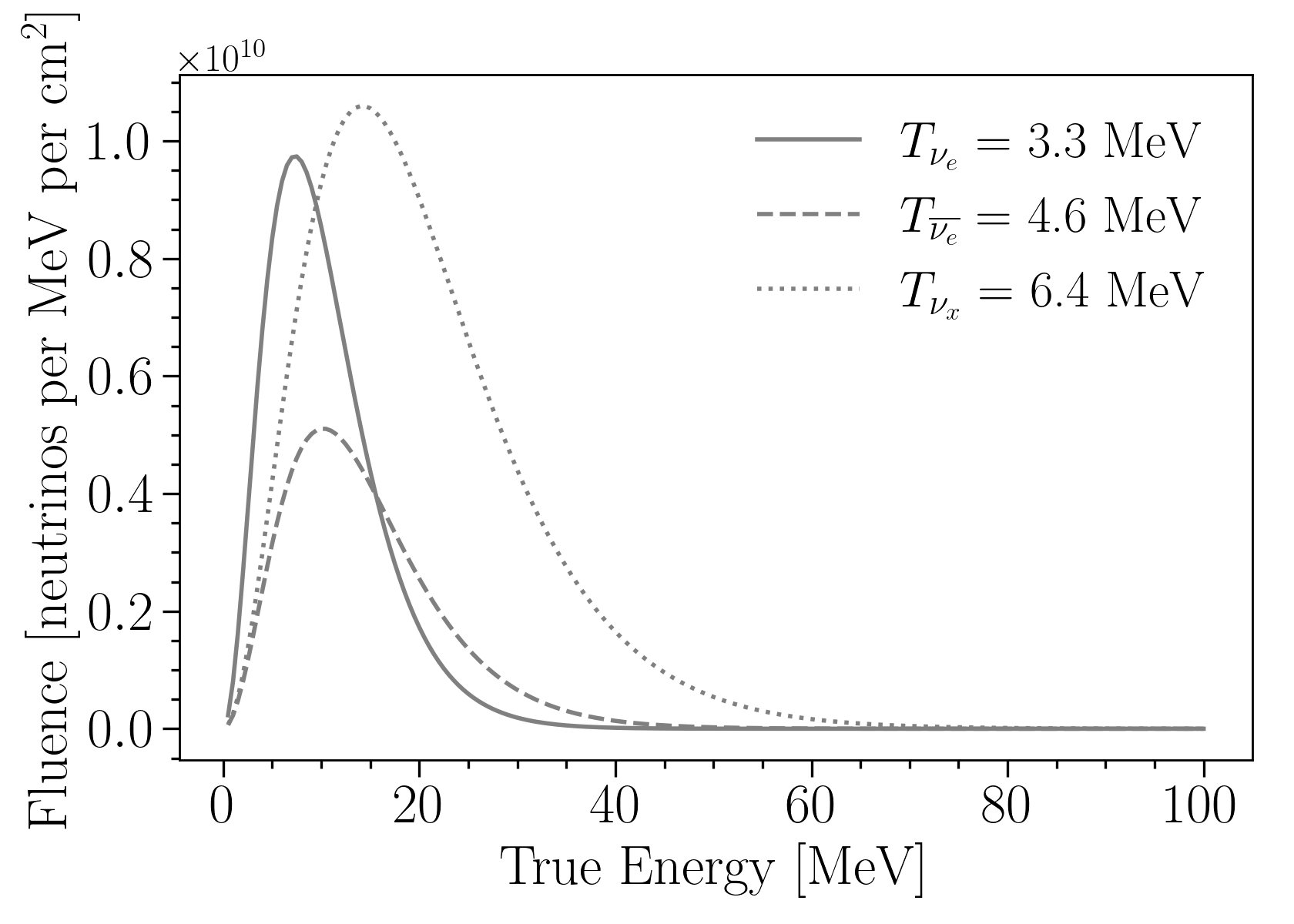}
 \caption{\textbf{Supernova neutrino differential fluence for a Fermi-Dirac spectrum.} The differential fluence is shown as a function of neutrino energy for each of the three neutrino flux components, assuming a Fermi-Dirac energy distribution, and the baseline supernova scenario described in~\ref{sec:methods}.}
  \label{fig:flux}
\end{figure}

We model the supernovae neutrino flux with a Fermi-Dirac spectrum, assuming a progenitor of 10 solar masses at a distance of \SI{10}{\kilo\parsec} from Earth as a benchmark.
The supernovae neutrino differential flux, as first described by in~\cite{Totani:1997vj}, is given by

\begin{equation}
\frac{dF_{\nu}}{d \varepsilon_{\nu}} = \frac{L_\nu}{4 \pi D^2 T^4_{\nu} F_3 (\eta)} \frac{\varepsilon_{\nu} ^2}{e^{\beta (\varepsilon_{\nu} - \mu ) } + 1},
\label{eq:flux}
\end{equation}
where $\varepsilon_\nu$ is incident neutrino energy, $L_{\nu}$ is the luminosity for each neutrino species of SN1987A integrated over total time of collapse for a total energy of approximately $3 \times 10^{53}$ ergs, $D$ is the distance from Earth to the supernova, $T_\nu$ is the average temperature of the neutrino species, $\mu$ is the chemical potential, $\beta = 1/ T_{\nu}$, and $\eta = \mu/T_{\nu}$ (the Boltzmann constant is set to unity). $F_3(\eta)$ is defined by 

\begin{equation}
F_3(\eta) \equiv \int_{0}^{\infty} \frac{x^n}{e^{x-\eta}+1} dx,
\end{equation}
and in this work $\eta$ is set to 0.
Fig~\ref{fig:flux} provides an example of this calculation, setting $T_{\nu_e} = \SI{3.3}\MeV$, $T_{\bar{\nu_e}} = \SI{4.6}\MeV$, and $T_{\nu_x} = \SI{6.4}\MeV$.

A pinching parameter can further modify the energy spectrum of supernova neutrinos to closer match the high energy tails of radiated neutrino spectra~\cite{Mirizzi:2015eza, Raffelt:2001kv, Raffelt:2003en, Buras:2002wt}. 
For measurements of the CC channel, where neutrino energy and observed energy are highly correlated, it is possible to resolve the two parameters the pinched spectrum model and even to differentiate between more complex energy spectra.

The characteristics of the NC channel, however, do not allow for such accurate measurements of the neutrino energy spectrum.
Incoherent NC interactions between \si{\MeV} scale neutrinos and Argon nuclei place the Argon nucleus into one of several excited states.
As the Argon nucleus returns to a relaxed state, it emits a $\gamma$-ray with discrete energy between $\SI{4.47}\MeV$ and $\SI{10.03}\MeV$, corresponding to the specific exited state.
While the threshold of these interactions is directly related to the excitation energy, the only other correlation between neutrino energy and observed energy comes from small differences in the cross-sections of the different excitation channels.
This decoupling of neutrino energy and observed energy results in limited sensitivity to the original neutrino energy spectrum.
For this reason, we choose a simple model, the Fermi-Dirac distribution, for the energy spectrum of each flux component.
While a pinching parameter can further modify the energy spectrum, this parameter gives changes to the flux that are similar in magnitude to changes from the average temperature, and can be degenerate with the average temperature~\cite{Minakata:2008nc}.
Therefore this analysis is concerned with using expected NC events to measure average temperature of the neutrino fluence without considering a pinching parameter.

\begin{figure}
  \includegraphics[width=\linewidth]{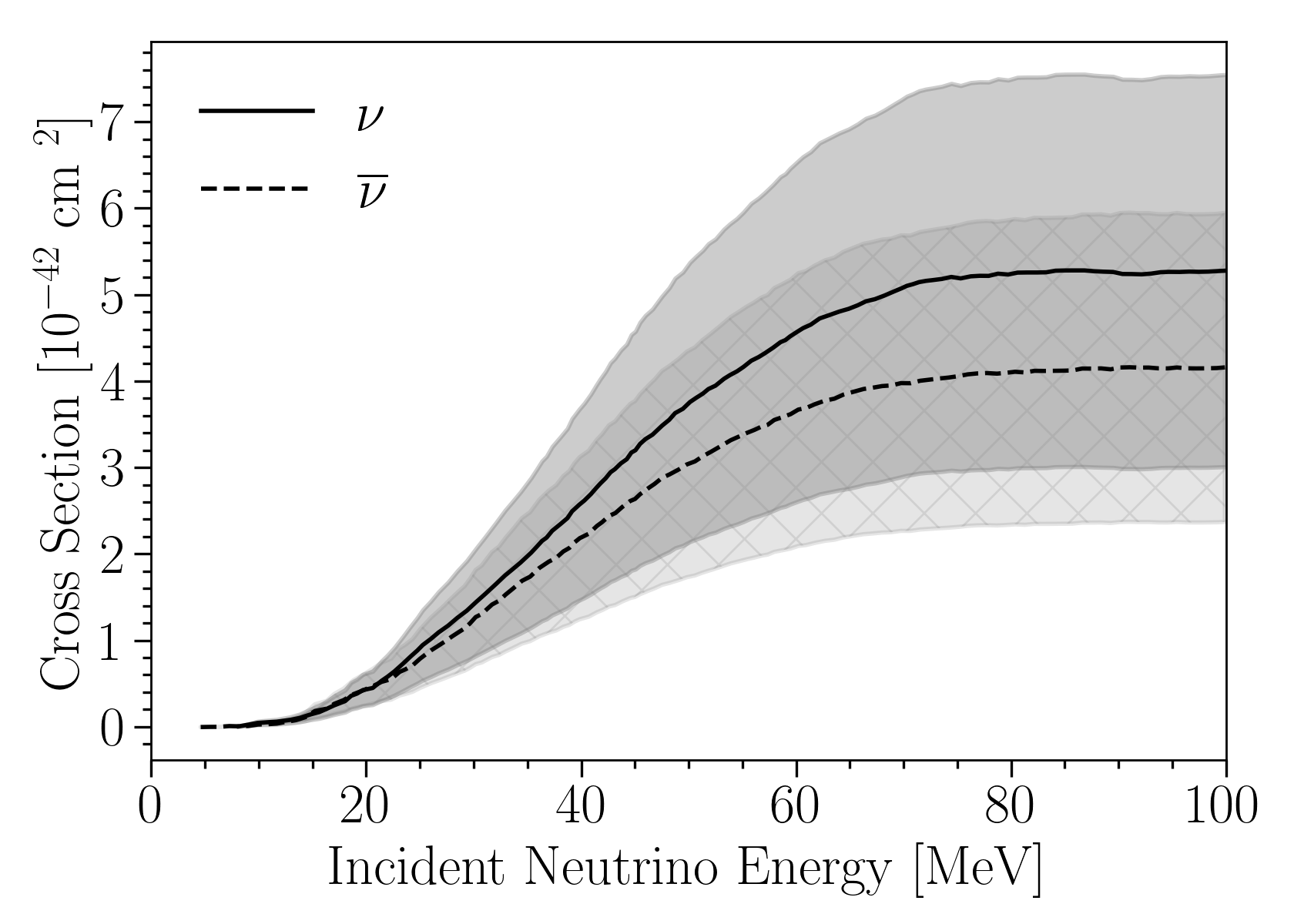}
  \caption{\textbf{Incoherent NC neutrino-Argon cross-sections and uncertainties.} The neutral-current neutrino-Argon cross-sections derived in~\cite{Tornow:2022kmo} are shown for $\nu$ and $\bar\nu$ as a function of incident neutrino energy with bands showing their associated ~43$\%$ uncertainties. The Solid shading $\nu$ cross-section uncertainties, and hatched shading shows the same for $\bar{\nu}$.}
    \label{fig:xsect}
\end{figure}

For neutrino interactions on Argon nuclei, we use the most recently published NC cross-sections from W. Tornow et. al.~\cite{Tornow:2022kmo}
Fig~\ref{fig:xsect} shows this cross-section for neutrinos and anti-neutrinos with $\SI{43}\percent$ theoretical uncertainty.
The uncertainty is composed of 15\% uncertainty on the B(M1$\uparrow$) measurements, which measure the transition strength of the $\gamma$-ray excitations and 40\% theoretical uncertainty due to differences derived from choice of shell-model calculation.

\begin{figure}
  \includegraphics[width=\linewidth]{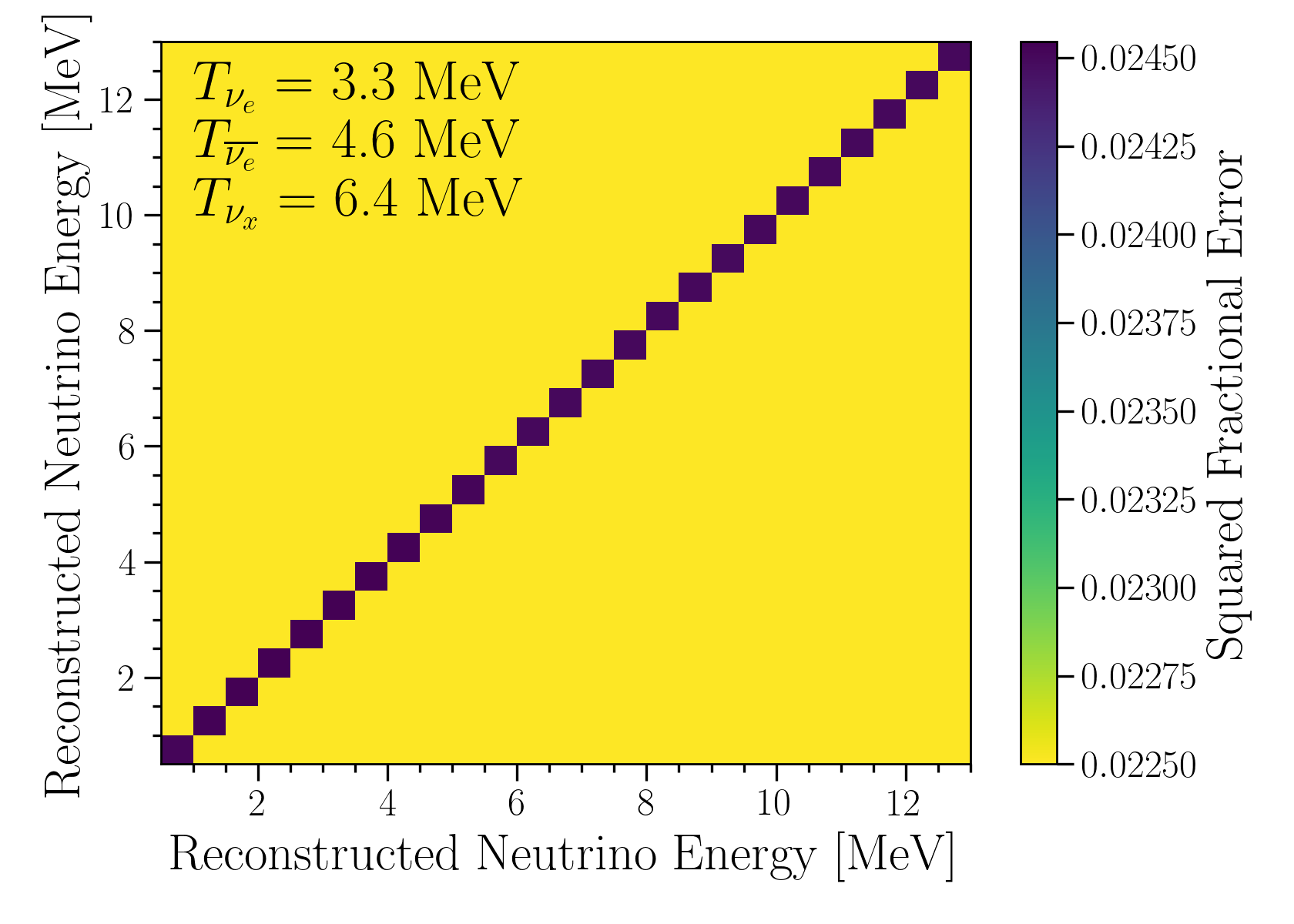}
  \caption{\textbf{DUNE FD NC fractional covariance matrix.} The uncertainties on the cross section are composed of 15\% correlated uncertainties and 40\% uncorrelated uncertainties, as described in~\cite{Tornow:2022kmo}}
  \label{fig:cov}
\end{figure}

We accounted for the uncertainty in the cross-section using a fractional covariance matrix comprised of 15\% correlated uncertainties within each excitation mode and 40\% uncorrelated uncertainties across the entire energy spectrum.
An example of the fractional covariance matrix is displayed in Fig~\ref{fig:cov}. For this analysis, the covariance matrix varies with input neutrino temperatures. To enable reconstruction of $\nu_x$ temperature across a wide range of injected temperatures, the fractional covariance matrices were computed at approximately every two \si\MeV{} of temperature and trilinear interpolation was utilized to generate the necessary fractional covariance matrix. 

\begin{figure}
  \includegraphics[width=\linewidth]{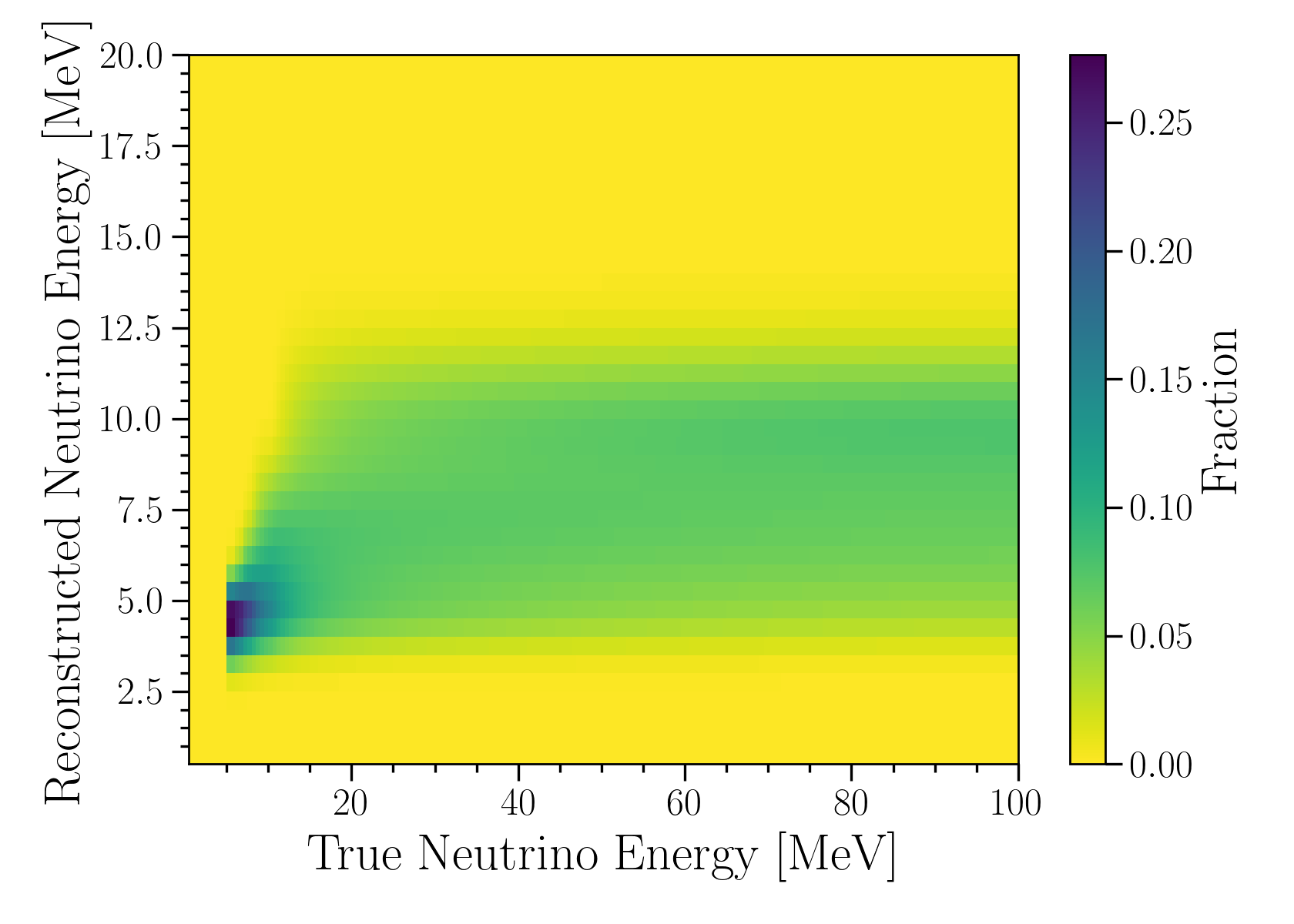}
  \caption{\textbf{DUNE FD NC response matrix.} Response matrix for the $\nu$ component of the incoherent NC neutrino interaction in LAr assuming 15\% energy resolution 
  for the DUNE FD TPC as described in~\cite{DUNE:2020ypp}.}
  \label{fig:smear}
\end{figure}

The final component necessary to model the supernovae signal in DUNE is the detector response and reconstruction efficiency.
DUNE expects 15\% energy resolution in their TPCs at the energies of relevance for supernovae neutrino measurements~\cite{DUNE:2020ypp}. 
Using this predicted energy resolution and the cross-sections of the individual photon excitation modes, we calculated a response matrix to capture the detector's response to NC events for the $\nu$ and $\bar{\nu}$ cross section channels.
This response matrix weights each excitation mode proportional to the interaction cross-section for each mode to produce a general response matrix, which we show in Fig.~\ref{fig:smear}.

\begin{figure}
  \includegraphics[width=\linewidth]{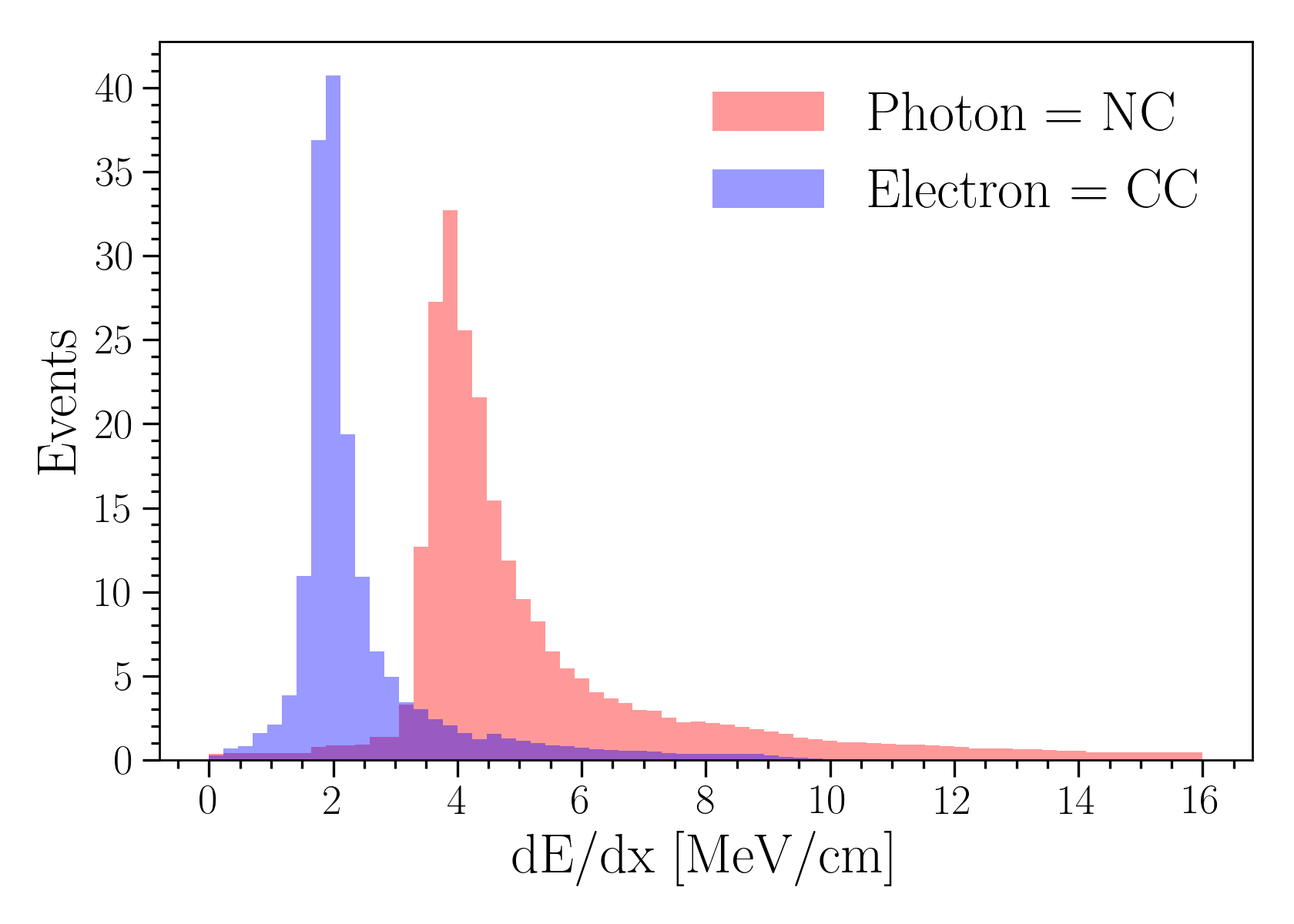}
  \caption{\textbf{DUNE dE/dx distribution for electron and photon like final states.} Distribution is normalized to the expected NC and CC event rates for a representative sample where $T_{\nu_x} = \SI{3.3} \MeV$, $T_{\bar{\nu_e}} = \SI{4.6} \MeV$, and $T_{\nu_x} = \SI{6.4} \MeV$.}
  \label{fig:dedx}
\end{figure}

We further explored efficiency cuts using the distribution of deposited energy per unit distance traveled in the detector to separate NC events from CC events. 
Using DUNE's expected dE/dx distribution for photon and electron particles, described in~\cite{DUNE:2020ypp}, we scaled these normalized distributions to the expected event rates in the NC and CC channels in our reconstructed energy region of interest between 0.5 and 13 \si\MeV{} for $T_{\nu_e} = \SI{3.3} \MeV$, $T_{\bar{\nu_e}} = \SI{4.6} \MeV$, and $T_{\nu_x} = \SI{6.4} \MeV$, shown in Fig.~\ref{fig:dedx}.

We found that one cut in the dE/dx region of 3.1 \si\MeV/cm includes 95.26\% of the signal events while rejecting all but 15.83\% of the background CC interactions due to the $\nu_e$ supernovae neutrino flux that DUNE will also be sensitive to. 

\begin{figure}
  \includegraphics[width=\linewidth]{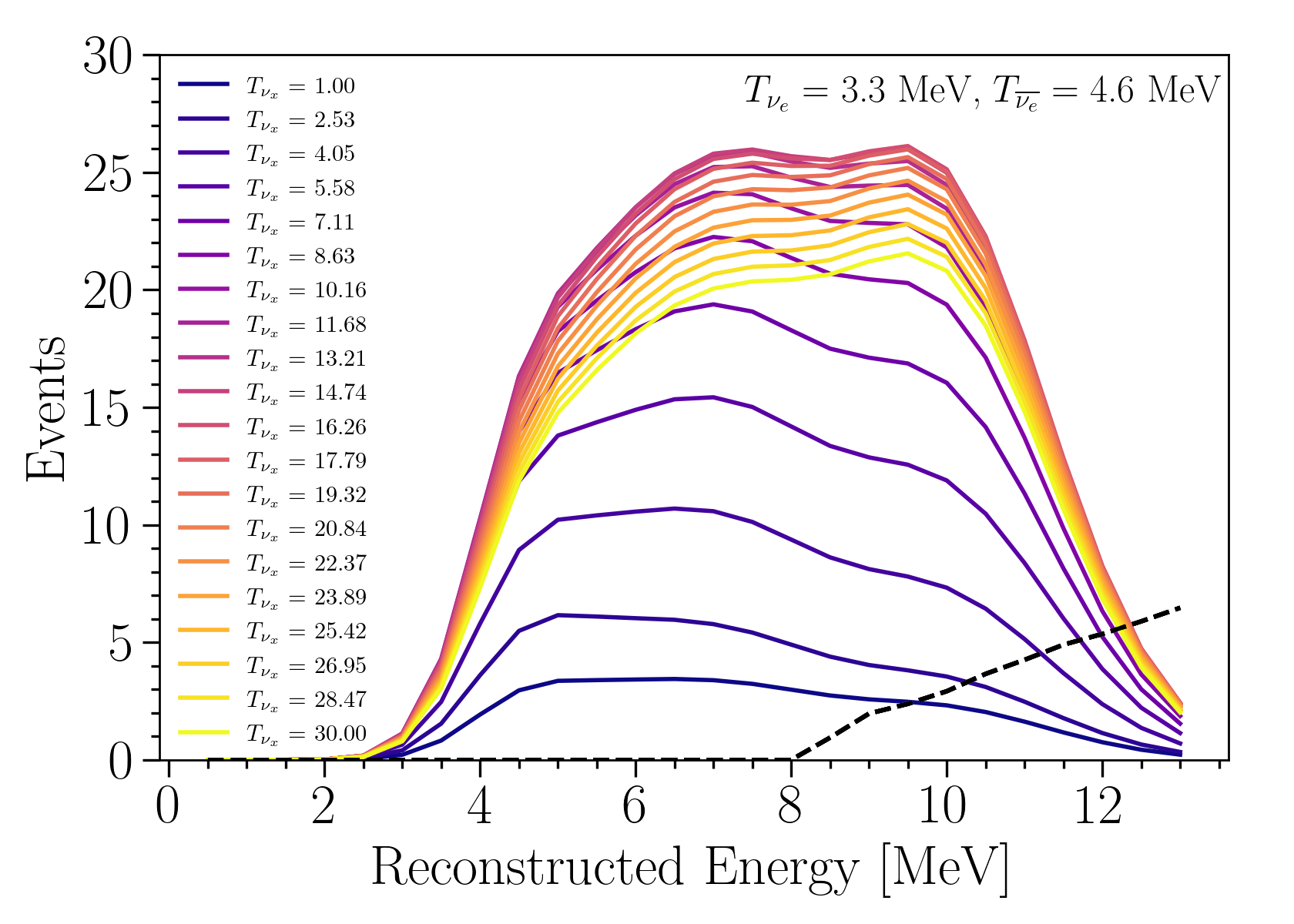}
  \caption{\textbf{Expected NC event rate in DUNE FD as a function of $\nu_x$ temperature}. $\nu_e$ and $\bar{\nu_e}$ temperatures are fixed to $\SI{3.3}\MeV$ and $\SI{4.6}\MeV$ respectively. The black dashed line represents events due to CC interactions in the DUNE FD from the $\nu_e$ component of the fluence.}
  \label{fig:event_rate}
\end{figure}

Fig.~\ref{fig:event_rate} is an example expected reconstructed event rate in DUNE FD, after combining the NC cross-section, response matrix, efficiency cut, and fluence for the specified neutrino temperatures.
This plots show the expected events for $T_{\nu_e} = \SI{3.3}\MeV$ and $T_{\bar{\nu_e}} = \SI{4.6}\MeV$ while average $T_{\nu_x}$ varies between $\SI{1}\MeV$ and \SI{30}\MeV. 
The dashed black line represents the CC events due to the $\nu_e$ component of the supernovae neutrino flux in DUNE. 
Although the NC channel is suppressed compared to the CC channel, DUNE's large detector size combined with the high fluence of expected neutrinos for a supernovae collapse provides us with a large enough sample to perform spectral analysis.

The goal of this analysis is to demonstrate reconstruction of the $\nu_x$ temperature by combining NC channel measurements from the DUNE FD and constraints on $\nu_e$ and $\bar{\nu_e}$ temperatures from CC measurements in DUNE and Hyper-K respectively. 
DUNE TPCs are predicted to resolve energies in our region of interest to around 15\%~\cite{DUNE:2020ypp}. Using this energy resolution to calculate the CC response matrix, $\nu_e$ fluence described in ~\ref{eq:flux}, and CC cross section~\cite{PhysRevC.80.055501}, we calculated the predicted CC event rate as a function of average $\nu_e$ temperature in DUNE and were able to reconstruct injected temperatures to 1 sigma credible region with \SI{0.5}\percent fractional error. 
We expect that Hyper-K will be able to reconstruct average $\bar{\nu_e}$ temperature to similar precision given their similar energy resolution~\cite{Bell:2020rkw} as the DUNE TPC, greater fiducial mass~\cite{Hyper-Kamiokande:2018ofw} than the DUNE FD modules, and greater cross-section for inverse beta decay events below $\sim\SI{15}\MeV$ than $\nu_e$ CC events on Argon~\cite{Scholberg:2012id}.

\section{Statistical methods\label{sec:stats}}
With the methods described above, we can now predict the NC event rates in DUNE for any combination of temperatures of the three neutrino flux components.
The NC data from DUNE provides information about the total neutrino flux, however, to obtain sensitivity to the $\nu_x$ temperature we must simultaneously constrain the temperature of the $\nu_e$ and $\bar{\nu_e}$ components.
We construct the binned-log-likelihood comparing expected NC DUNE data to our predictions, and combine this with external statistical constraints on $\nu_e$ and $\bar{\nu_e}$.
This combination produces a constrained log-likelihood given by

\begin{equation}
\mathcal{L}(\vec{T})=\left[\prod_i\frac{\lambda_i(\vec{T},\alpha_i)^{k_i} e^{-\lambda_i(\vec{T},\alpha_i)}}{k_i!}\right] \mathcal{P}\left(T_{\nu_e}\right) \mathcal{P}\left(T_{\bar{\nu_e}}\right) \mathcal{P}(\vec\alpha),
\label{eq:like}
\end{equation}
where $\lambda_i(\vec{T},\alpha_i)$ is the expected number of events in bin $i$ for the combination of temperatures $\vec{T}$, and $k_i$ is the number of observed data events in bin $i$.
For the two external constraints, $\mathcal{P}(T_{\nu_e})$ and $\mathcal{P}(T_{\bar{\nu_e}})$, we assume Normally distributed constraints centered on the true values of the temperatures with widths that correspond to the uncertainties on $\nu_e$ and $\bar{\nu_e}$ temperatures that we expect to obtain from the CC measurements of DUNE and Hyper-K, assumed to be $\SI{0.5}\percent$.
To model the cross-section uncertainties, we introduce the parameters $\vec\alpha$ which fractionally modify the expected number of events in each bin, such that $\lambda_i(\vec{T},\alpha_i)=(1+\alpha_i)\cdot\lambda_i(\vec{T})$.
The parameters $\vec\alpha$ are constrained by the multivariate normal distribution denoted by $\mathcal{P}(\vec{\alpha})$, which has a fractional covariance matrix derived from the cross-section uncertainties.

To explore the sensitivity of this combined analysis to the $\nu_x$ average temperature, we examine a set of representative scenarios where $T_{\nu_e}=\SI{3.3}\MeV$, $T_{\bar{\nu_e}}=\SI{4.6}\MeV$, and $T_{\nu_x}$ has a value between $\SI{0.5}\MeV$ and $\SI{30}\MeV$.
For each value of $T_{\nu_x}$ we perform an Asimov test, where the nominal expected event distribution is injected as data
We use a Markov Chain Monte Carlo with adaptive parallel tempering~\cite{Foreman-Mackey:2012any, Vousden_2015} to explore the parameter space, and derive Bayesian credible regions (CR) for $T_{\nu_x}$ that are marginalized over $T_{\nu_e}$, $T_{\bar{\nu_e}}$, and the nuisance parameters $\vec\alpha$.
The two electron flavor temperatures have normal priors modeled after the expected constraints from CC data, and the nuisance parameters $\vec\alpha$ have a multivariate-normal prior, as described in~\ref{eq:like}.
We use a uniform prior for $T_{\nu_x}$ across most of the parameter space, but introduce a hyperbolic tangent cutoff at $\SI{60}\MeV$, with a characteristic transition width of $\SI{3}\MeV$.
For injected $T_{\nu_x}$ less than $\SI{10}\MeV$ a degenerate region of the parameter space becomes apparent, which extends to reconstructed temperatures above $\SI{100}\MeV$ for the lowest injected temperatures.
We focus the results presented here on temperatures below $\SI{60}\MeV$, because such large temperatures are not within the expected range, and will be ruled out by other observations~\cite{Beacom:2010kk,Woosley:1988ip,Keil:2002in,Yoshida:2005uy,Heger:2003mm,Yuksel:2005ae}.
This is accomplished with the hyperbolic tangent cutoff, the primary effect of which is a reduction in the size of the derived CR's for injected $T_{\nu_x}$ less than $\SI{5}\MeV$, because there is no longer appreciable posterior mass above $\SI{60}\MeV$.
Additional discussion of this prior, and CR's derived with a uniform $T_{\nu_x}$ prior from $\SI{0.1}\MeV$ to $\SI{300}\MeV$ are given in Appendix~\ref{sec:appendix}.

\section{Results}
The average $\nu_e$ and $\bar{\nu_e}$ temperatures will be constrained very well by CC measurements from DUNE and Hyper-K.
As a result, the sensitivity to $T_{\nu_x}$ does not significantly depend on the choice of injected $T_{\nu_e}$ and $T_{\bar{\nu_e}}$.
We can therefore examine the sensitivity to reconstructed $\nu_x$ temperature as a function of true injected $\nu_x$ temperature without significant bias from the choice of injected $\nu_e$ and $\bar{\nu_e}$ temperatures.
We present sensitivities to $T_{\nu_x}$, using $T_{\nu_e}=\SI{3.3}{\MeV}$ and $T_{\nu_e}=\SI{4.6}{\MeV}$ as a representative set of temperatures.

\begin{figure}
  \includegraphics[width=\linewidth]{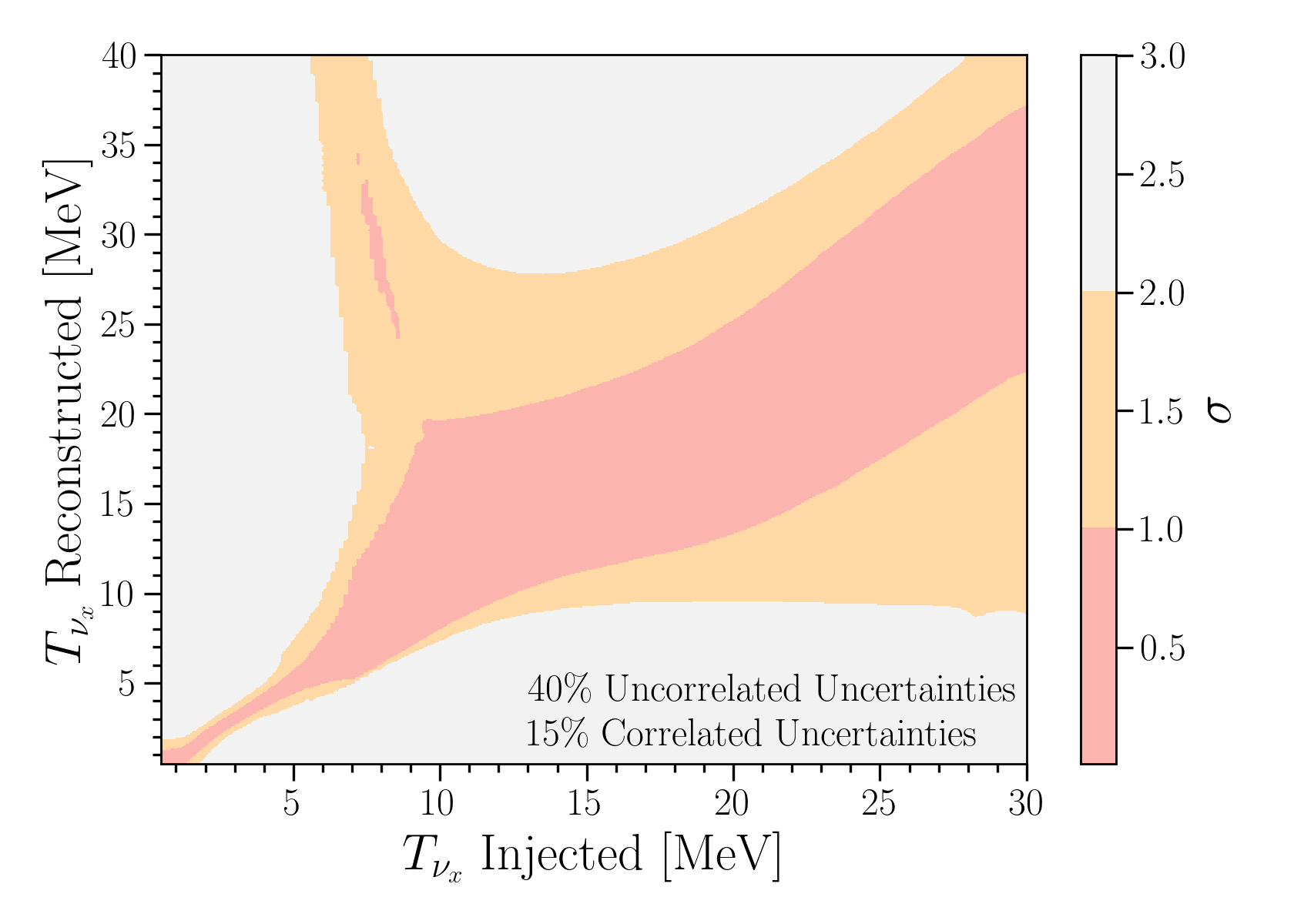}
  \caption{\textbf{Expected $T_{\nu_x}$ credible regions with full cross-section uncertainties.} The Asimov credible regions for $T_{\nu_x}$ are shown as a function of injected $T_{\nu_x}$ by the colored regions. Average neutrino temperatures of $T_{\nu_e}=\SI{3.3} \MeV$ and $T_{\bar{{\nu_e}}}=\SI{4.6} \MeV$ are assumed, along with 40\% uncorrelated uncertainties and 15\% correlated uncertainties on the NC cross-section.}
  \label{fig:full_uncer}
\end{figure}

\begin{figure}
  \includegraphics[width=\linewidth]{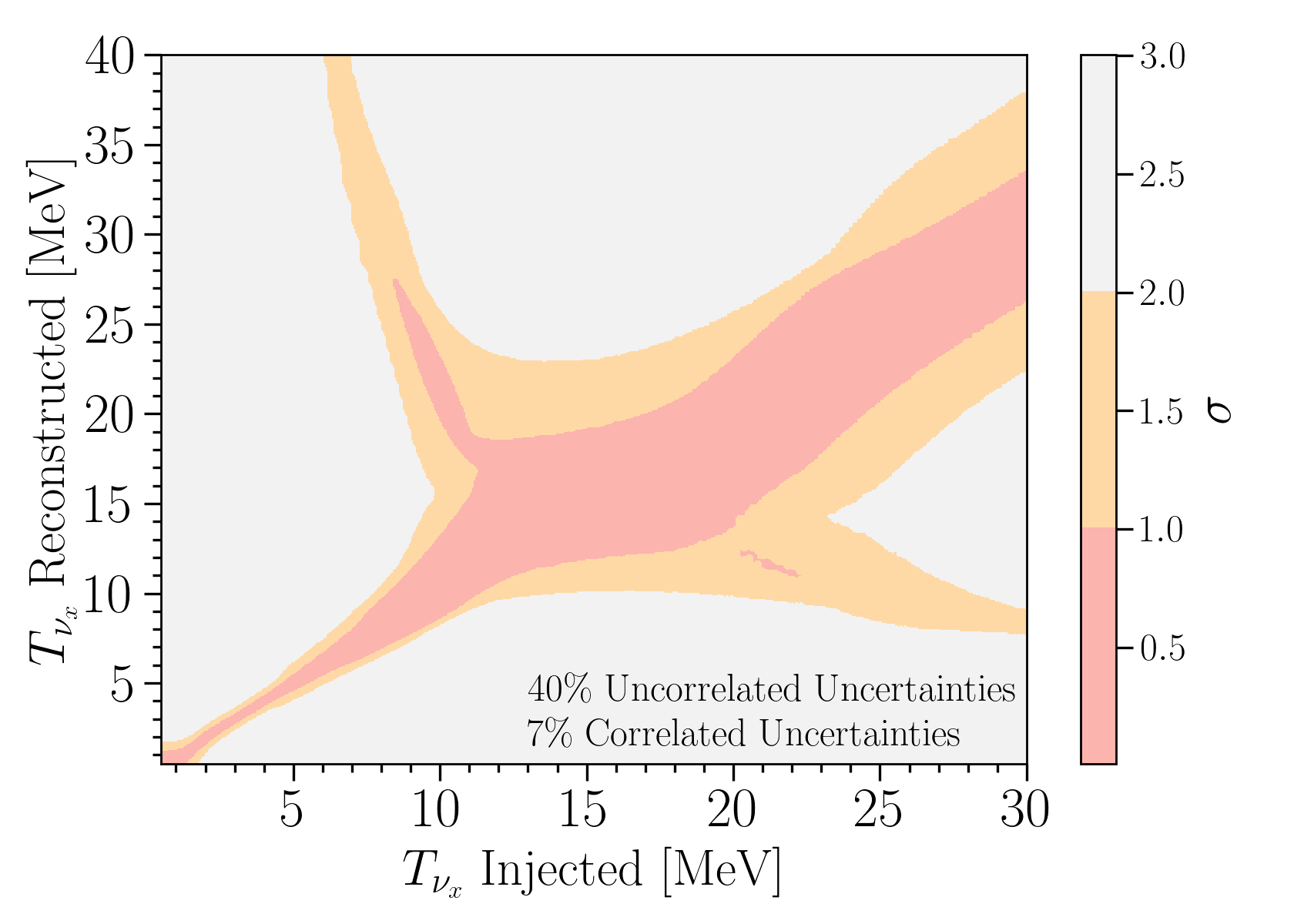}
  \caption{\textbf{Expected $T_{\nu_x}$ credible regions with reduced cross-section uncertainties.} The Asimov credible regions for $T_{\nu_x}$ are shown as a function of injected $T_{\nu_x}$ by the colored regions. Average neutrino temperatures of $T_{\nu_e}=\SI{3.3} \MeV$ and $T_{\bar{{\nu_e}}}=\SI{4.6} \MeV$ are assumed, along with 40\% uncorrelated uncertainties and 7\% correlated uncertainties on the NC cross-section.}
  \label{fig:halved correlated}
\end{figure}

\begin{figure}
  \includegraphics[width=\linewidth]{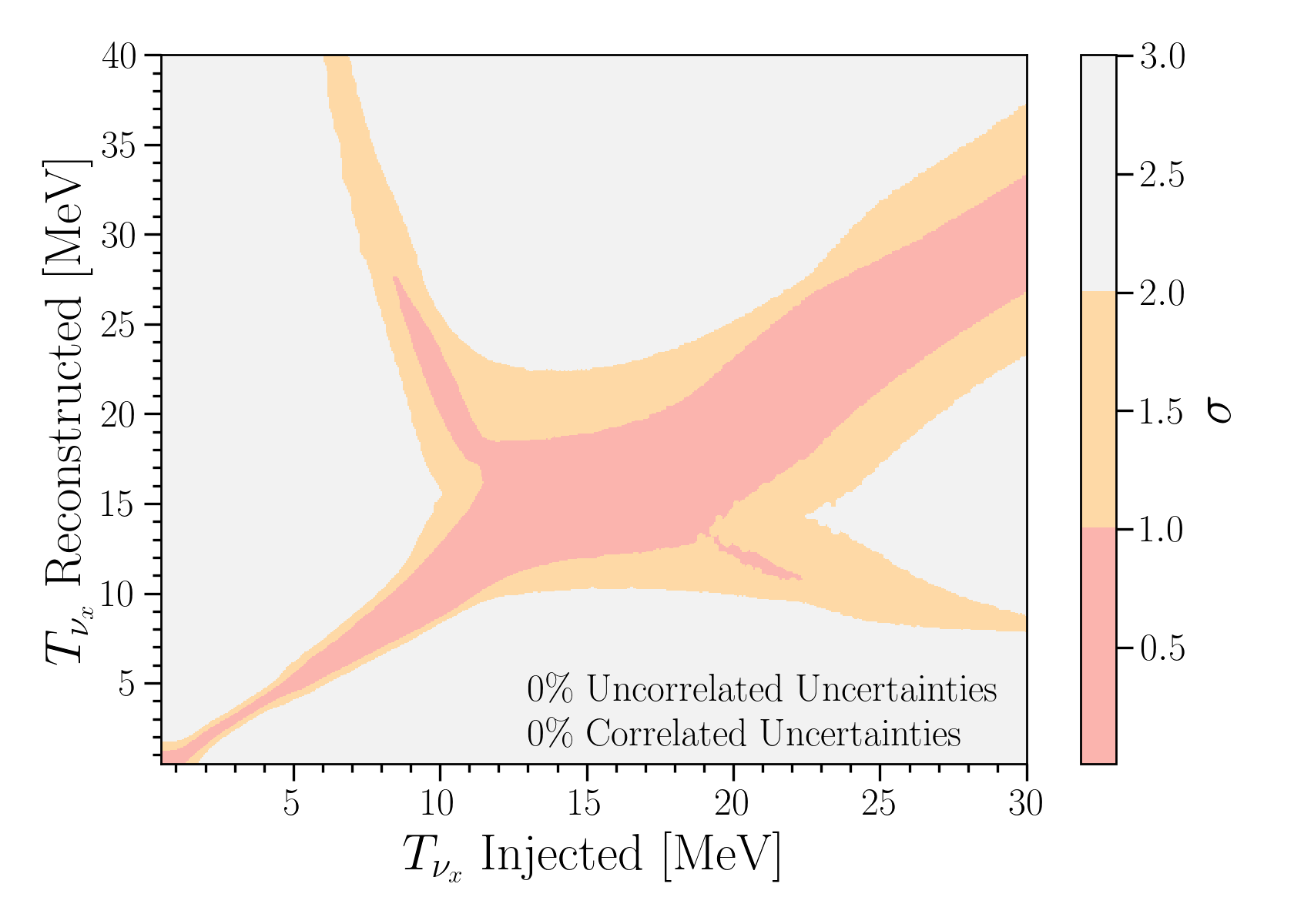}
  \caption{\textbf{Expected $T_{\nu_x}$ credible regions with no cross-section uncertainties.} The Asimov credible regions for $T_{\nu_x}$ are shown as a function of injected $T_{\nu_x}$ by the colored regions. Average neutrino temperatures of $T_{\nu_e}=\SI{3.3} \MeV$ and $T_{\bar{{\nu_e}}}=\SI{4.6} \MeV$ are assumed, along no uncertainties on the NC cross-section.}
  \label{fig:no_uncer}
\end{figure}

We explore sensitivities for  three scenarios with different cross-section uncertainties: a case without cross-section uncertainties, a case with full cross-section uncertainties, and a case with reduced cross-section uncertainties.
The case with full cross-section uncertainties assumes $\SI{40}\percent$ fully uncorrelated uncertainty in the cross-section and a $\SI{15}\percent$ correlated uncertainty in the cross section as described in~\ref{sec:methods}.
The case with reduced cross-section uncertainties retains the $\SI{40}\percent$ fully uncorrelated uncertainty, but reduces the correlated uncertainty to $\SI{7}\percent$ to model the effect of improved B(M1$\uparrow$) measurements.
Fig.~\ref{fig:no_uncer}, ~\ref{fig:full_uncer}, and ~\ref{fig:halved correlated} show the expected sensitivity to $T_{\nu_x}$ as Asimov credible regions for the no uncertainties, full uncertainties, and reduced uncertainties cases respectively.

There are two distinct degenerate regions that arise in the measurement space from multiple factors.
The primary cause is the discrete $\gamma$-ray energies of the NC cross-section that results in a decoupling between neutrino energy and observed energy.
In addition to that, for $T_{\nu_x}$ above approximately $\SI{8}\MeV$ the total fluence becomes much more uniform across the energy range of the NC excitations.
These characteristics of the NC cross-section combined with the Fermi-Dirac spectrum at higher average temperatures leads to a degeneracy in the $\nu_x$ temperature measurement.

\begin{figure}
  \includegraphics[width=\linewidth]{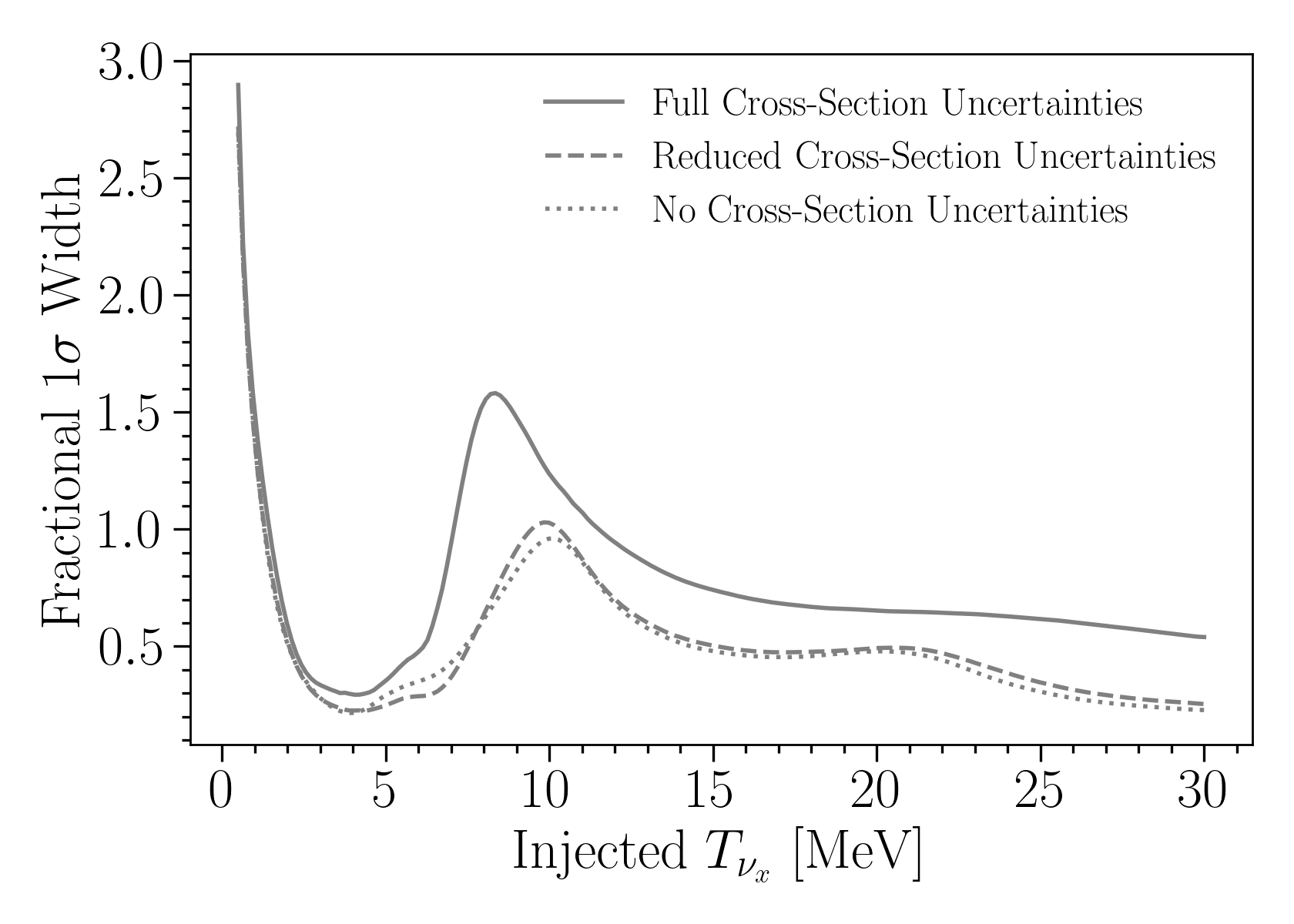}
  \caption{\textbf{Fractional width of the 1$\sigma$ credible region.} Solid line represents the scenario with 40\% uncorrelated uncertainties and 15\% correlated uncertainties on the NC cross-section. Dashed line represents the scenario with 40\% uncorrelated uncertainties and 7\% correlated uncertainties on the NC cross-section. Dotted line represents the scenario with no uncertainties on the NC cross-section.}
  \label{fig:frx_width}
\end{figure}

Fig.~\ref{fig:frx_width} shows the fractional width of the 1$\sigma$ credible region for these measurements.
The peak in the fractional width of the 1$\sigma$ credible region around $\SI{10} \MeV$ true $\nu_x$ temperature is due to the degeneracy in the measurement. 
Reducing the correlated uncertainties from 15\% to 7\%, without changes to the 40\% uncorrelated uncertainties, reduces the fractional width of the 1$\sigma$ credible region almost to that of the measurement without any uncertainties on the cross section.
This demonstrates that even modest improvements to the Argon B(M1$\uparrow$) measurements provide substantial improvements to the $T_{\nu_x}$ sensitivity.

\section{Conclusions}
While NC channel measurements are often overlooked due to the technical challenges and small sample sizes, this analysis has shown the power of NC measurements in measuring $\nu_x$ spectral information from Type-II core collapse supernovae events. 
Due to the discrete excitation nature of this interaction channel, the energy spectrum from supernovae neutrinos cannot be resolved, but general parameters of the fluence can be measured to the 1$\sigma$ and 2$\sigma$ level. 
For this analysis we focused on measuring the average $\nu_x$ temperature from the incoherent NC neutrino-Argon interaction in the DUNE FD modules.
To make this measurement we combined the expected incoherent NC sample from the DUNE FD with expected constraints from the CC channel measurements of the DUNE and Hyper-K experiments on $T{\nu_e}$ and $T_{\bar\nu_e}$ of the supernovae flux.
We find that such a combined analysis can measure the average $\nu_x$ temperature to within a factor of two in most cases, and to within 30\% in the best case.

In addition to exploring DUNE's sensitivity to average $\nu_x$ temperature using the NC channel, we explore the effects of reducing the uncertainty in NC cross-section predictions.
By reducing correlated uncertainties from 15\% to 7\%, the $T_{\nu_x}$ measurement becomes sample size limited, and approaches the case where there are no uncertainties on the cross-section.
More precise measurements of B(M1$\uparrow$) transition strengths will provide even greater spectral information for the $\nu_x$ component of supernovae flux.
While this measurement will be limited by small energy range of Argon-nucleus gamma emissions from NC interactions, combining NC measurements of neutrinos on Argon with NC measurements of neutrinos on Carbon from JUNO could provide stronger constraints on average $\nu_x$ temperature reconstruction. 

\section{Acknowledgements\label{sec:acknowledgements}}
We want to thank Janet M. Conrad for insights on liquid Argon TPCs. We would also like to thank Anna C. Hayes for her dedicated work on modeling neutral current cross-sections. DAN is supported by the NSF Graduate Research Fellowship under Grant No. 2141064. AS is supported by the U.S. Department of Energy through the Los Alamos National Laboratory. Los Alamos National Laboratory is operated by Triad National Security, LLC, for the National Nuclear Security Administration of U.S. Department of Energy (Contract No. 89233218CNA000001).

\bibliographystyle{apsrev}
\bibliography{apssamp} 

\begin{thebibliography}{24}
\expandafter\ifx\csname natexlab\endcsname\relax\def\natexlab#1{#1}\fi
\expandafter\ifx\csname bibnamefont\endcsname\relax
  \def\bibnamefont#1{#1}\fi
\expandafter\ifx\csname bibfnamefont\endcsname\relax
  \def\bibfnamefont#1{#1}\fi
\expandafter\ifx\csname citenamefont\endcsname\relax
  \def\citenamefont#1{#1}\fi
\expandafter\ifx\csname url\endcsname\relax
  \def\url#1{\texttt{#1}}\fi
\expandafter\ifx\csname urlprefix\endcsname\relax\def\urlprefix{URL }\fi
\providecommand{\bibinfo}[2]{#2}
\providecommand{\eprint}[2][]{\url{#2}}

\bibitem[{\citenamefont{Mirizzi et~al.}(2016)\citenamefont{Mirizzi, Tamborra,
  Janka, Saviano, Scholberg, Bollig, Hudepohl, and
  Chakraborty}}]{Mirizzi:2015eza}
\bibinfo{author}{\bibfnamefont{A.}~\bibnamefont{Mirizzi}},
  \bibinfo{author}{\bibfnamefont{I.}~\bibnamefont{Tamborra}},
  \bibinfo{author}{\bibfnamefont{H.-T.} \bibnamefont{Janka}},
  \bibinfo{author}{\bibfnamefont{N.}~\bibnamefont{Saviano}},
  \bibinfo{author}{\bibfnamefont{K.}~\bibnamefont{Scholberg}},
  \bibinfo{author}{\bibfnamefont{R.}~\bibnamefont{Bollig}},
  \bibinfo{author}{\bibfnamefont{L.}~\bibnamefont{Hudepohl}}, \bibnamefont{and}
  \bibinfo{author}{\bibfnamefont{S.}~\bibnamefont{Chakraborty}},
  \bibinfo{journal}{Riv. Nuovo Cim.} \textbf{\bibinfo{volume}{39}},
  \bibinfo{pages}{1} (\bibinfo{year}{2016}), \eprint{1508.00785}.

\bibitem[{\citenamefont{Abud et~al.}(2023)}]{duneCC}
\bibinfo{author}{\bibfnamefont{A.~A.} \bibnamefont{Abud}} \bibnamefont{et~al.}
  (\bibinfo{collaboration}{DUNE Collaboration}) (\bibinfo{year}{2023}),
  \eprint{2303.17007}.

\bibitem[{\citenamefont{Abe et~al.}(2021)}]{Hyper-Kamiokande:2021frf}
\bibinfo{author}{\bibfnamefont{K.}~\bibnamefont{Abe}} \bibnamefont{et~al.}
  (\bibinfo{collaboration}{Hyper-Kamiokande}), \bibinfo{journal}{Astrophys. J.}
  \textbf{\bibinfo{volume}{916}}, \bibinfo{pages}{15} (\bibinfo{year}{2021}),
  \eprint{2101.05269}.

\bibitem[{\citenamefont{Raffelt}(2001)}]{Raffelt:2001kv}
\bibinfo{author}{\bibfnamefont{G.~G.} \bibnamefont{Raffelt}},
  \bibinfo{journal}{Astrophys. J.} \textbf{\bibinfo{volume}{561}},
  \bibinfo{pages}{890} (\bibinfo{year}{2001}), \eprint{astro-ph/0105250}.

\bibitem[{\citenamefont{Raffelt et~al.}(2003)\citenamefont{Raffelt, Keil,
  Buras, Janka, and Rampp}}]{Raffelt:2003en}
\bibinfo{author}{\bibfnamefont{G.~G.} \bibnamefont{Raffelt}},
  \bibinfo{author}{\bibfnamefont{M.~T.} \bibnamefont{Keil}},
  \bibinfo{author}{\bibfnamefont{R.}~\bibnamefont{Buras}},
  \bibinfo{author}{\bibfnamefont{H.-T.} \bibnamefont{Janka}}, \bibnamefont{and}
  \bibinfo{author}{\bibfnamefont{M.}~\bibnamefont{Rampp}}, in
  \emph{\bibinfo{booktitle}{{4th Workshop on Neutrino Oscillations and their
  Origin (NOON2003)}}} (\bibinfo{year}{2003}), pp. \bibinfo{pages}{380--387},
  \eprint{astro-ph/0303226}.

\bibitem[{\citenamefont{Buras et~al.}(2003)\citenamefont{Buras, Janka, Keil,
  Raffelt, and Rampp}}]{Buras:2002wt}
\bibinfo{author}{\bibfnamefont{R.}~\bibnamefont{Buras}},
  \bibinfo{author}{\bibfnamefont{H.-T.} \bibnamefont{Janka}},
  \bibinfo{author}{\bibfnamefont{M.~T.} \bibnamefont{Keil}},
  \bibinfo{author}{\bibfnamefont{G.~G.} \bibnamefont{Raffelt}},
  \bibnamefont{and} \bibinfo{author}{\bibfnamefont{M.}~\bibnamefont{Rampp}},
  \bibinfo{journal}{Astrophys. J.} \textbf{\bibinfo{volume}{587}},
  \bibinfo{pages}{320} (\bibinfo{year}{2003}), \eprint{astro-ph/0205006}.

\bibitem[{\citenamefont{Tornow et~al.}(2022)\citenamefont{Tornow, Tonchev,
  Finch, Krishichayan, Wang, Hayes, Yeomans, and Newmark}}]{Tornow:2022kmo}
\bibinfo{author}{\bibfnamefont{W.}~\bibnamefont{Tornow}},
  \bibinfo{author}{\bibfnamefont{A.~P.} \bibnamefont{Tonchev}},
  \bibinfo{author}{\bibfnamefont{S.~W.} \bibnamefont{Finch}},
  \bibinfo{author}{\bibnamefont{Krishichayan}},
  \bibinfo{author}{\bibfnamefont{X.~B.} \bibnamefont{Wang}},
  \bibinfo{author}{\bibfnamefont{A.~C.} \bibnamefont{Hayes}},
  \bibinfo{author}{\bibfnamefont{H.~G.~D.} \bibnamefont{Yeomans}},
  \bibnamefont{and} \bibinfo{author}{\bibfnamefont{D.~A.}
  \bibnamefont{Newmark}}, \bibinfo{journal}{Phys. Lett. B}
  \textbf{\bibinfo{volume}{835}}, \bibinfo{pages}{137576}
  (\bibinfo{year}{2022}), \eprint{2210.14316}.

\bibitem[{\citenamefont{Bhattacharya et~al.}(2009)\citenamefont{Bhattacharya,
  Goodman, and Garc\'{\i}a}}]{PhysRevC.80.055501}
\bibinfo{author}{\bibfnamefont{M.}~\bibnamefont{Bhattacharya}},
  \bibinfo{author}{\bibfnamefont{C.~D.} \bibnamefont{Goodman}},
  \bibnamefont{and}
  \bibinfo{author}{\bibfnamefont{A.}~\bibnamefont{Garc\'{\i}a}},
  \bibinfo{journal}{Phys. Rev. C} \textbf{\bibinfo{volume}{80}},
  \bibinfo{pages}{055501} (\bibinfo{year}{2009}),
  \urlprefix\url{https://link.aps.org/doi/10.1103/PhysRevC.80.055501}.

\bibitem[{\citenamefont{Abi et~al.}(2020)}]{DUNE:2020ypp}
\bibinfo{author}{\bibfnamefont{B.}~\bibnamefont{Abi}} \bibnamefont{et~al.}
  (\bibinfo{collaboration}{DUNE}) (\bibinfo{year}{2020}), \eprint{2002.03005}.

\bibitem[{\citenamefont{Cavanna}(2022)}]{flavio}
\bibinfo{author}{\bibfnamefont{F.}~\bibnamefont{Cavanna}},
  \bibinfo{howpublished}{Personal communication} (\bibinfo{year}{2022}).

\bibitem[{\citenamefont{Djurcic et~al.}(2015)}]{JUNO:2015sjr}
\bibinfo{author}{\bibfnamefont{Z.}~\bibnamefont{Djurcic}} \bibnamefont{et~al.}
  (\bibinfo{collaboration}{JUNO}) (\bibinfo{year}{2015}), \eprint{1508.07166}.

\bibitem[{\citenamefont{Totani et~al.}(1998)\citenamefont{Totani, Sato, Dalhed,
  and Wilson}}]{Totani:1997vj}
\bibinfo{author}{\bibfnamefont{T.}~\bibnamefont{Totani}},
  \bibinfo{author}{\bibfnamefont{K.}~\bibnamefont{Sato}},
  \bibinfo{author}{\bibfnamefont{H.~E.} \bibnamefont{Dalhed}},
  \bibnamefont{and} \bibinfo{author}{\bibfnamefont{J.~R.}
  \bibnamefont{Wilson}}, \bibinfo{journal}{Astrophys. J.}
  \textbf{\bibinfo{volume}{496}}, \bibinfo{pages}{216} (\bibinfo{year}{1998}),
  \eprint{astro-ph/9710203}.

\bibitem[{\citenamefont{Minakata et~al.}(2008)\citenamefont{Minakata, Nunokawa,
  Tomas, and Valle}}]{Minakata:2008nc}
\bibinfo{author}{\bibfnamefont{H.}~\bibnamefont{Minakata}},
  \bibinfo{author}{\bibfnamefont{H.}~\bibnamefont{Nunokawa}},
  \bibinfo{author}{\bibfnamefont{R.}~\bibnamefont{Tomas}}, \bibnamefont{and}
  \bibinfo{author}{\bibfnamefont{J.~W.~F.} \bibnamefont{Valle}},
  \bibinfo{journal}{JCAP} \textbf{\bibinfo{volume}{12}}, \bibinfo{pages}{006}
  (\bibinfo{year}{2008}), \eprint{0802.1489}.

\bibitem[{\citenamefont{Bell et~al.}(2020)\citenamefont{Bell, Dolan, and
  Robles}}]{Bell:2020rkw}
\bibinfo{author}{\bibfnamefont{N.~F.} \bibnamefont{Bell}},
  \bibinfo{author}{\bibfnamefont{M.~J.} \bibnamefont{Dolan}}, \bibnamefont{and}
  \bibinfo{author}{\bibfnamefont{S.}~\bibnamefont{Robles}},
  \bibinfo{journal}{JCAP} \textbf{\bibinfo{volume}{09}}, \bibinfo{pages}{019}
  (\bibinfo{year}{2020}), \eprint{2005.01950}.

\bibitem[{\citenamefont{Abe et~al.}(2018)}]{Hyper-Kamiokande:2018ofw}
\bibinfo{author}{\bibfnamefont{K.}~\bibnamefont{Abe}} \bibnamefont{et~al.}
  (\bibinfo{collaboration}{Hyper-Kamiokande}) (\bibinfo{year}{2018}),
  \eprint{1805.04163}.

\bibitem[{\citenamefont{Scholberg}(2012)}]{Scholberg:2012id}
\bibinfo{author}{\bibfnamefont{K.}~\bibnamefont{Scholberg}},
  \bibinfo{journal}{Ann. Rev. Nucl. Part. Sci.} \textbf{\bibinfo{volume}{62}},
  \bibinfo{pages}{81} (\bibinfo{year}{2012}), \eprint{1205.6003}.

\bibitem[{\citenamefont{Foreman-Mackey
  et~al.}(2013)\citenamefont{Foreman-Mackey, Hogg, Lang, and
  Goodman}}]{Foreman-Mackey:2012any}
\bibinfo{author}{\bibfnamefont{D.}~\bibnamefont{Foreman-Mackey}},
  \bibinfo{author}{\bibfnamefont{D.~W.} \bibnamefont{Hogg}},
  \bibinfo{author}{\bibfnamefont{D.}~\bibnamefont{Lang}}, \bibnamefont{and}
  \bibinfo{author}{\bibfnamefont{J.}~\bibnamefont{Goodman}},
  \bibinfo{journal}{Publ. Astron. Soc. Pac.} \textbf{\bibinfo{volume}{125}},
  \bibinfo{pages}{306} (\bibinfo{year}{2013}), \eprint{1202.3665}.

\bibitem[{\citenamefont{Vousden et~al.}(2015)\citenamefont{Vousden, Farr, and
  Mandel}}]{Vousden_2015}
\bibinfo{author}{\bibfnamefont{W.~D.} \bibnamefont{Vousden}},
  \bibinfo{author}{\bibfnamefont{W.~M.} \bibnamefont{Farr}}, \bibnamefont{and}
  \bibinfo{author}{\bibfnamefont{I.}~\bibnamefont{Mandel}},
  \bibinfo{journal}{Monthly Notices of the Royal Astronomical Society}
  \textbf{\bibinfo{volume}{455}}, \bibinfo{pages}{1919} (\bibinfo{year}{2015}),
  \urlprefix\url{https://doi.org/10.1093/mnras/stv2422}.

\bibitem[{\citenamefont{Beacom}(2010)}]{Beacom:2010kk}
\bibinfo{author}{\bibfnamefont{J.~F.} \bibnamefont{Beacom}},
  \bibinfo{journal}{Ann. Rev. Nucl. Part. Sci.} \textbf{\bibinfo{volume}{60}},
  \bibinfo{pages}{439} (\bibinfo{year}{2010}), \eprint{1004.3311}.

\bibitem[{\citenamefont{Woosley and Haxton}(1988)}]{Woosley:1988ip}
\bibinfo{author}{\bibfnamefont{S.~E.} \bibnamefont{Woosley}} \bibnamefont{and}
  \bibinfo{author}{\bibfnamefont{W.~C.} \bibnamefont{Haxton}},
  \bibinfo{journal}{Nature} \textbf{\bibinfo{volume}{334}}, \bibinfo{pages}{45}
  (\bibinfo{year}{1988}).

\bibitem[{\citenamefont{Keil et~al.}(2003)\citenamefont{Keil, Raffelt, and
  Janka}}]{Keil:2002in}
\bibinfo{author}{\bibfnamefont{M.~T.} \bibnamefont{Keil}},
  \bibinfo{author}{\bibfnamefont{G.~G.} \bibnamefont{Raffelt}},
  \bibnamefont{and} \bibinfo{author}{\bibfnamefont{H.-T.} \bibnamefont{Janka}},
  \bibinfo{journal}{Astrophys. J.} \textbf{\bibinfo{volume}{590}},
  \bibinfo{pages}{971} (\bibinfo{year}{2003}), \eprint{astro-ph/0208035}.

\bibitem[{\citenamefont{Yoshida et~al.}(2005)\citenamefont{Yoshida, Kajino, and
  Hartmann}}]{Yoshida:2005uy}
\bibinfo{author}{\bibfnamefont{T.}~\bibnamefont{Yoshida}},
  \bibinfo{author}{\bibfnamefont{T.}~\bibnamefont{Kajino}}, \bibnamefont{and}
  \bibinfo{author}{\bibfnamefont{D.~H.} \bibnamefont{Hartmann}},
  \bibinfo{journal}{Phys. Rev. Lett.} \textbf{\bibinfo{volume}{94}},
  \bibinfo{pages}{231101} (\bibinfo{year}{2005}), \eprint{astro-ph/0505043}.

\bibitem[{\citenamefont{Heger et~al.}(2005)\citenamefont{Heger, Kolbe, Haxton,
  Langanke, Martinez-Pinedo, and Woosley}}]{Heger:2003mm}
\bibinfo{author}{\bibfnamefont{A.}~\bibnamefont{Heger}},
  \bibinfo{author}{\bibfnamefont{E.}~\bibnamefont{Kolbe}},
  \bibinfo{author}{\bibfnamefont{W.~C.} \bibnamefont{Haxton}},
  \bibinfo{author}{\bibfnamefont{K.}~\bibnamefont{Langanke}},
  \bibinfo{author}{\bibfnamefont{G.}~\bibnamefont{Martinez-Pinedo}},
  \bibnamefont{and} \bibinfo{author}{\bibfnamefont{S.~E.}
  \bibnamefont{Woosley}}, \bibinfo{journal}{Phys. Lett. B}
  \textbf{\bibinfo{volume}{606}}, \bibinfo{pages}{258} (\bibinfo{year}{2005}),
  \eprint{astro-ph/0307546}.

\bibitem[{\citenamefont{Yuksel et~al.}(2006)\citenamefont{Yuksel, Ando, and
  Beacom}}]{Yuksel:2005ae}
\bibinfo{author}{\bibfnamefont{H.}~\bibnamefont{Yuksel}},
  \bibinfo{author}{\bibfnamefont{S.}~\bibnamefont{Ando}}, \bibnamefont{and}
  \bibinfo{author}{\bibfnamefont{J.~F.} \bibnamefont{Beacom}},
  \bibinfo{journal}{Phys. Rev. C} \textbf{\bibinfo{volume}{74}},
  \bibinfo{pages}{015803} (\bibinfo{year}{2006}), \eprint{astro-ph/0509297}.

\end{thebibliography}

\newpage\hbox{}\thispagestyle{empty}\newpage

\widetext
\appendix
\section{$T_{\nu_x}$ Prior\label{sec:appendix}} 

In this work we demonstrate the potential sensitivity of the DUNE FD to the average temperature of the supernova $\nu_x$ component with credible regions derived from the highest posterior density regions (HPD).
A side effect of this methodology is a dependence of the local HPD width on the global distribution of posterior mass.
For these NC derived constraints on $T_{\nu_x}$, a degenerate region of allowed $\nu_x$ temperatures is present for the entire energy range.
Below $\sim\SI{10}\MeV$ and above $\sim\SI{20}\MeV$, this degeneracy manifests as two distinct allowed regions at both the $1\sigma$ and $2\sigma$ level.
For some values of injected $T_{\nu_x}$ the allowed region not centered on the injected temperature, referred to as the degenerate region, lies within the range of expected supernova neutrino temperatures.
However, for injected $T_{\nu_x}$ less than $\sim\SI{5}\MeV$ this degenerate region exists at temperatures far above the expected range of supernova neutrino temperatures.
In the main text of this paper, we choose to apply a prior that allows only values of $T_{\nu_x}$ below $\SI{60}\MeV$.
This is motivated by the low temperatures observed from SN1987A~\cite{Woosley:1988ip}, predicted by supernova simulation studies~\cite{Keil:2002in}, required by constraints from neutrino induced nucleosynthesis~\cite{Yoshida:2005uy,Heger:2003mm}, and disfavored by diffuse supernova background searches~\cite{Yuksel:2005ae}.
This results in expected sensitivities that are not biased by a degeneracy that is both far outside of the range of expected supernova behavior and which would be ruled out by other measurements.

\begin{figure*}
     \subfloat[40\% uncorrelated uncertainties and 15\% correlated uncertainties on the NC cross-section]{
         \includegraphics[width=0.4\linewidth]{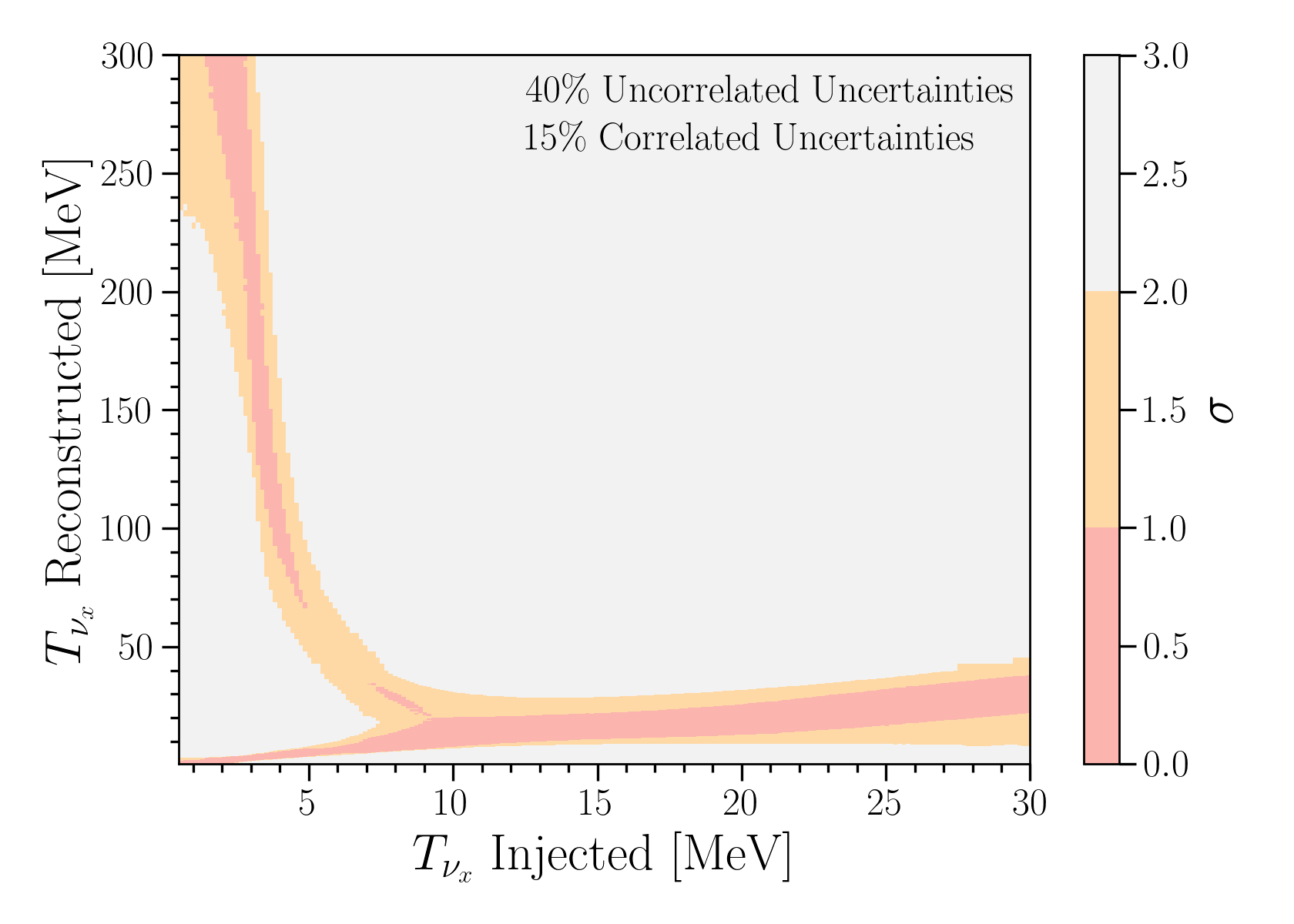}}
    \subfloat[40\% uncorrelated uncertainties and 7\% correlated uncertainties on the NC cross-section]{
         \includegraphics[width=0.4\linewidth]{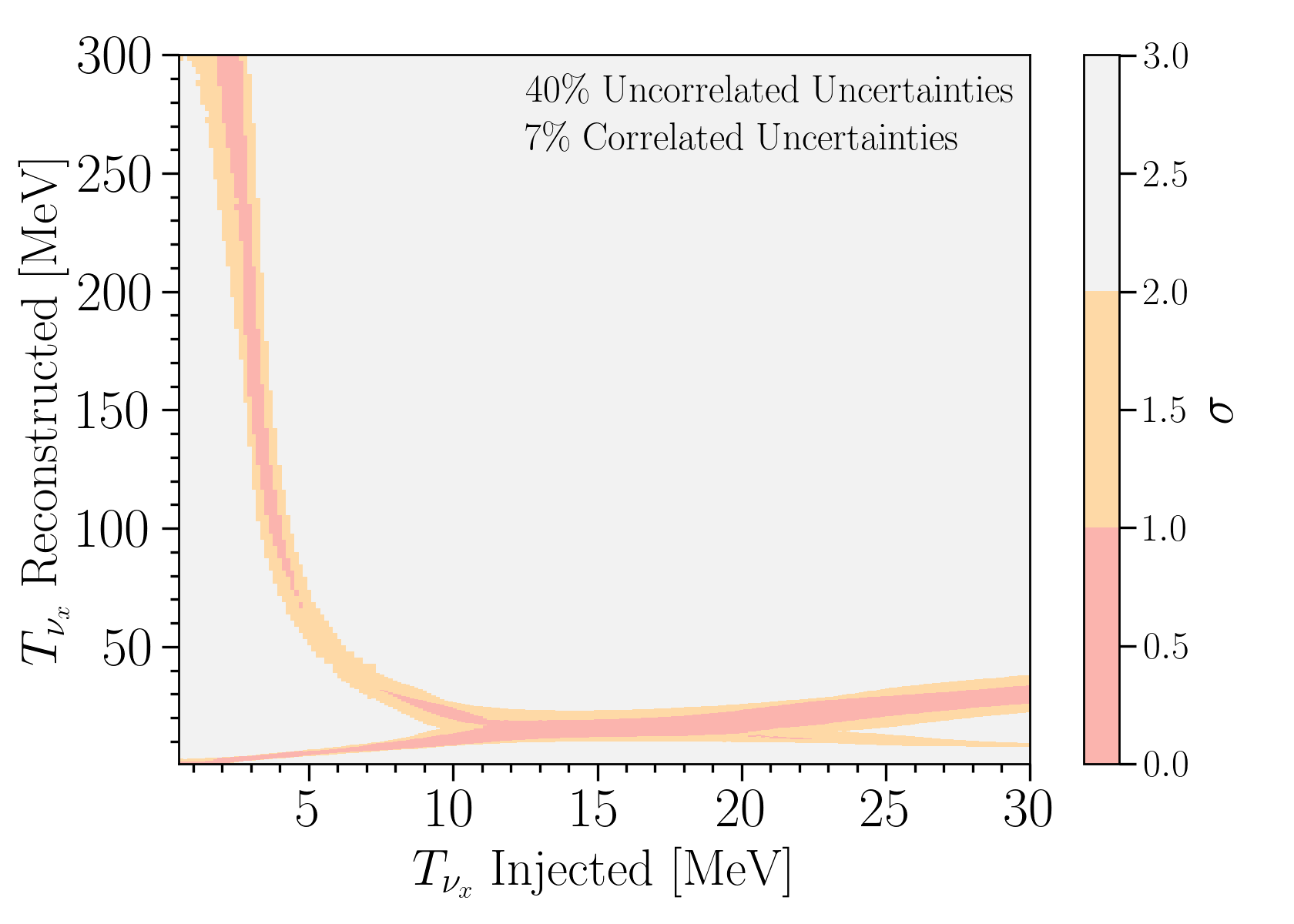}}
    \\
    \subfloat[No uncertainties on the NC cross-section]{
         \includegraphics[width=0.4\linewidth]{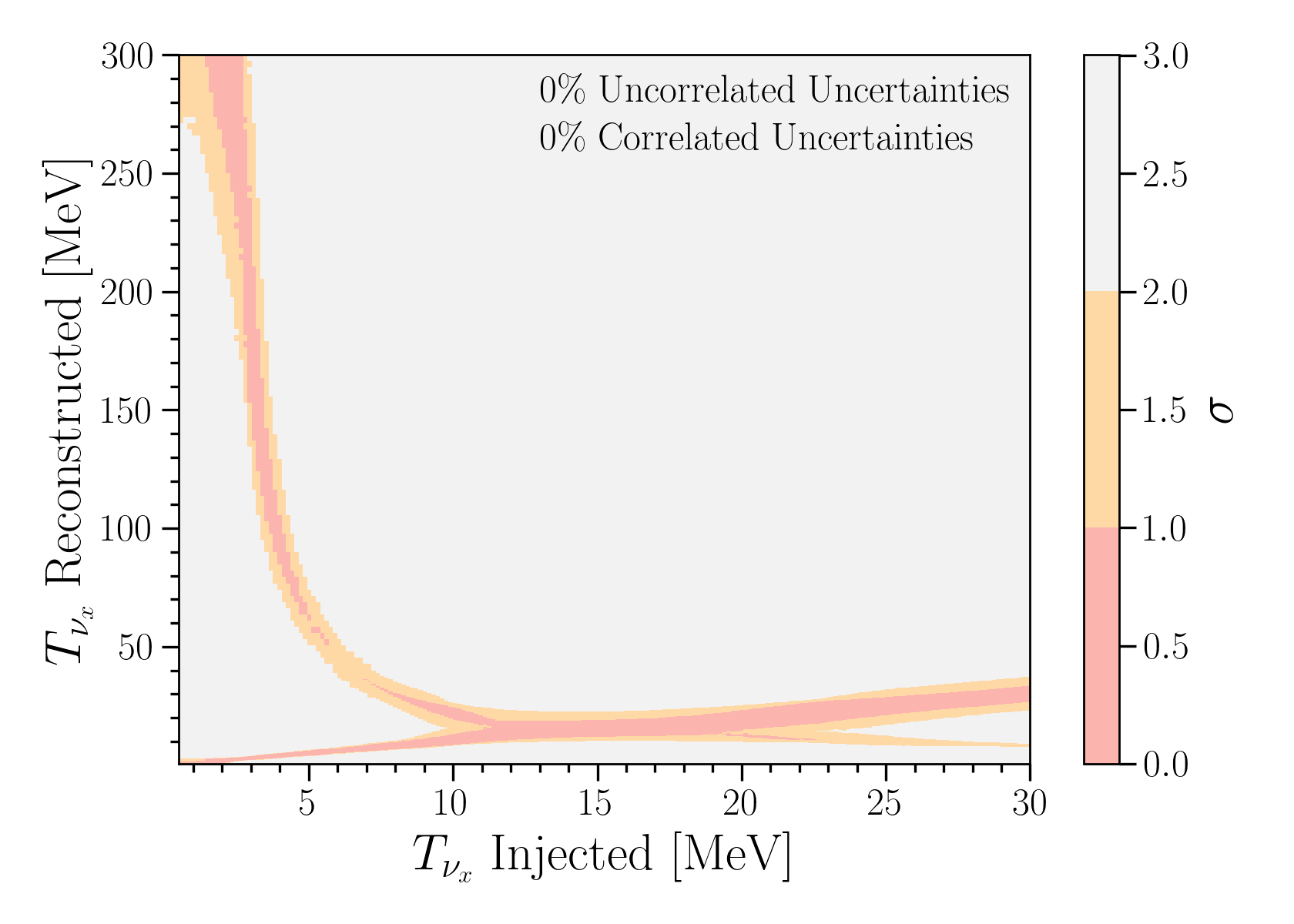}}
    \caption{\textbf{Expected $T_{\nu_x}$ credible regions with a uniform prior from $\SI{0.1}{\MeV}$ to $\SI{300}{\MeV}$.} The Asimov credible regions for $T_{\nu_x}$ are shown as a function of injected $T_{\nu_x}$ by the colored regions. Average neutrino temperatures of $T_{\nu_e}=\SI{3.3} \MeV$ and $T_{\bar{{\nu_e}}}=\SI{4.6} \MeV$ are assumed. The credible regions in this figure are computed with a uniform $T_{\nu_x}$ prior that extends up $\SI{300}\MeV$; this is in contrast to the credible regions shown in Figures~\ref{fig:full_uncer},~\ref{fig:halved correlated},~\ref{fig:no_uncer} which limited values of $T_{\nu_x}$ to be below $\SI{60}\MeV$.}
     \label{fig:appendix_measurements}
\end{figure*}

In this appendix we show the expected sensitivities under a different prior assumption, namely a uniform prior on $T_{\nu_x}$ between $\SI{0.1}\MeV$ and $\SI{300}\MeV$.
Because the degenerate region extends above $\SI{60}\MeV$ at low $\nu_x$ temperatures and contains appreciable posterior mass, this has the effect of widening the allowed regions of $T_{\nu_x}$ that are centered on the injected $T_{\nu_x}$.
Fig.~\ref{fig:appendix_measurements} show the expected sensitivity to $T_{\nu_x}$ with this wider range of allowed $T_{\nu_x}$ values for the three different cross-section uncertainty scenarios explored in the main text.
The degenerate region below $\sim\SI{5}\MeV$ extends up into the $100$'s of $\si\MeV$, and below $\sim\SI{3}\MeV$ the degenerate region begins to intersect with the prior boundary at $\SI{300}\MeV$.

\begin{figure}[t]
  \includegraphics[width=0.5
\linewidth]{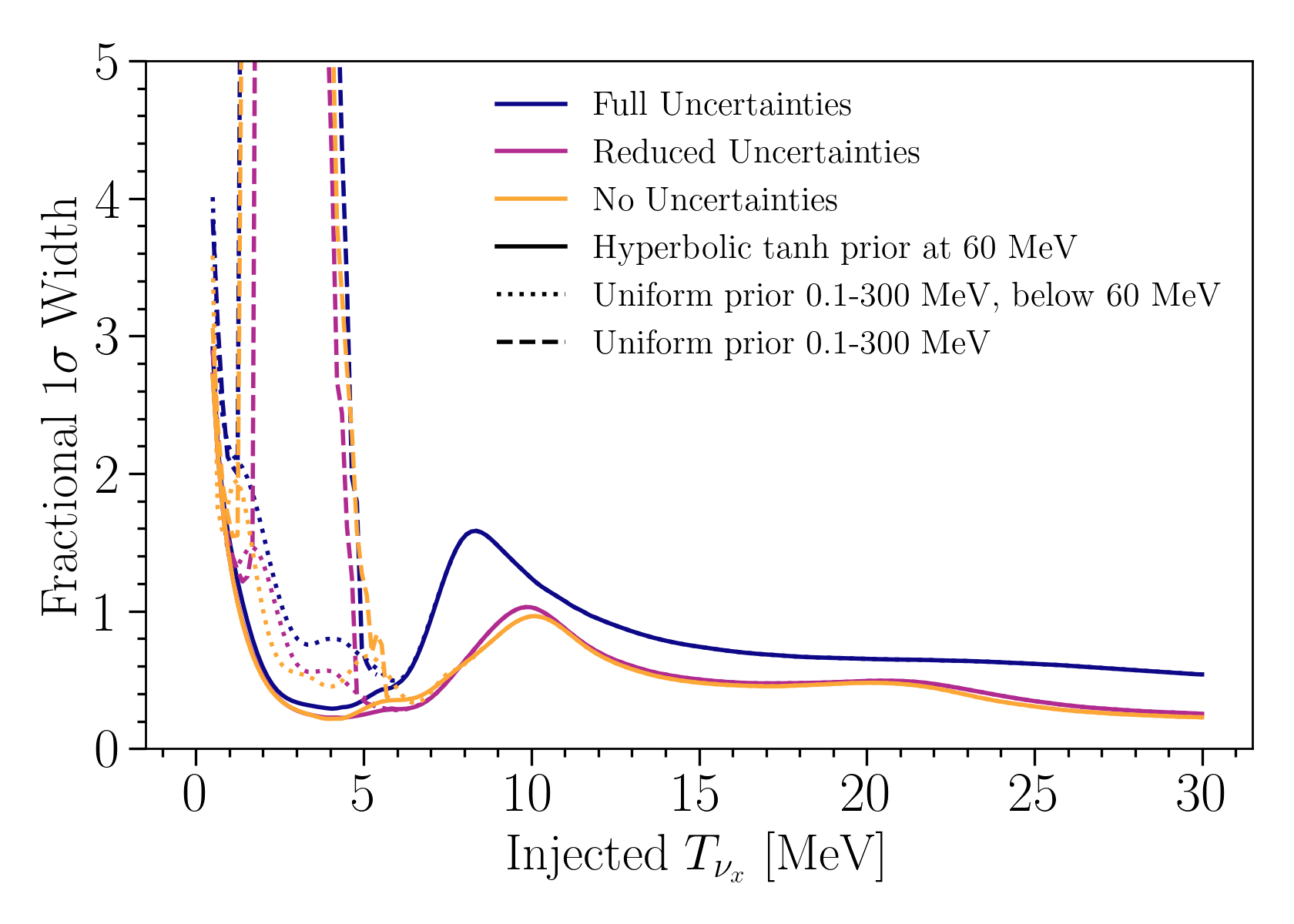}
  \caption{\textbf{Fractional width of $1\sigma$ credible regions.} The solid lines correspond to the width from the hyperbolic tangent prior, the dotted lines correspond to the width below $\SI{60}\MeV$ from the $0.1-300~\si\MeV$ uniform prior, and the dashed lines correspond to the full width from the from the $0.1-300~\si\MeV$ uniform prior (which extends up to 35). The three colors correspond to the three cases of cross section uncertainties.}
  \label{fig:width_all}
\end{figure}

To examine the effect of this prior we compare the width of the $1\sigma$ credible region across three scenarios: the full width derived from a uniform $T_{\nu_x}$ prior between $\SI{0.1}\MeV$ and $\SI{300}\MeV$, the width of the allowed region below $\SI{60}\MeV$ using the same uniform $T_{\nu_x}$ prior between $\SI{0.1}\MeV$ and $\SI{300}\MeV$, and finally the width derived using the prior from the main text (a hyperbolic-tangent cutoff prior that penalizes $T_{\nu_x}$ above $\SI{60}\MeV$ and has a $\SI{3}\MeV$ characteristic width).
Fig.~\ref{fig:width_all} shows the fractional width of these $1\sigma$ credible regions for the three scenarios.
The solid lines denote the case where $T_{\nu_x}$ prior is in the form of a hyperbolic tangent cutoff at $\SI{60}\MeV$.
Relaxing the prior assumptions on $T_{\nu_x}$ significantly widens the allowed regions below $\SI{60}\MeV$ for injected $T_{\nu_x}$ between $\SI{1}\MeV$ and $\SI{5}\MeV$, but does not significantly effect the expected allowed regions outside of this region.
The width of the allowed regions considering the full $T_{\nu_x}$ parameter space up to $\SI{300}\MeV$ remains largely unchanged outside of this $\SI{1}\MeV$ to $\SI{5}\MeV$ region of injected $T_{\nu_x}$, but between $\SI{1}\MeV$ and $\SI{5}\MeV$ is dominated by the large width of the high-temperature degenerate region.

\end{document}